\documentclass[graybox]{svmult}

\usepackage{type1cm}        % activate if the above 3 fonts are
\usepackage{makeidx}         % allows index generation
\usepackage{graphicx}        % standard LaTeX graphics tool
\usepackage{multicol}        % used for the two-column index
\usepackage[bottom]{footmisc}% places footnotes at page bottom

\usepackage[colorlinks=true, urlcolor= blue ,linkcolor=blue]{hyperref}

\usepackage{newtxtext}       %
\usepackage[varvw]{newtxmath}       % selects Times Roman as basic font

\usepackage{etoolbox}

\makeatletter
\newenvironment{sqcases}{%
    \matrix@check\sqcases\env@sqcases
}{%
    \endarray\right.%
}
\def\env@sqcases{%
    \let\@ifnextchar\new@ifnextchar
    \left\lbrack
    \def\arraystretch{1.2}%
    \array{@{}l@{\quad}l@{}}%
}
\makeatother

\makeindex             % used for the subject index
\begin{document}

\title*{Towards $\mathcal{N}=2$ higher-spin supergravity}
\author{
    Ioseph Buchbinder\orcidID{0000-0002-7968-901X},
    \\
    Evgeny Ivanov\orcidID{0000-0002-5594-2826}
 and\\
Nikita Zaigraev\orcidID{0000-0002-4385-8723}}
\institute{Ioseph Buchbinder \at Bogoliubov Laboratory of Theoretical Physics, JINR,
    141980 Dubna, Moscow region, Russia, \email{buchbinder@theor.jinr.ru},
\and
    Evgeny Ivanov \at Bogoliubov Laboratory of Theoretical Physics, JINR,
    141980 Dubna, Moscow region, Russia,
    \\
    Moscow Institute of Physics and Technology,
    141700 Dolgoprudny, Moscow region, Russia,
    \\ \email{eivanov@theor.jinr.ru}
\and Nikita Zaigraev \at Bogoliubov Laboratory of Theoretical Physics, JINR,
141980 Dubna, Moscow region, Russia,
\\
Moscow Institute of Physics and Technology,
141700 Dolgoprudny, Moscow region, Russia,
\\
 \email{nikita.zaigraev@phystech.edu}}
\maketitle

\textit{To the blessed memory of Professor A.A. Starobinsky, outstanding scientist and remarkable person}

\hspace{2mm}

\abstract*{Each chapter should be preceded by an abstract (no more than 200 words) that summarizes the content.
The abstract will appear \textit{online} at \url{www.SpringerLink.com} and be available with unrestricted access. This allows unregistered users to read the abstract as a teaser for the complete chapter.
Please use the 'starred' version of the \texttt{abstract} command for typesetting the text of the online abstracts (cf. source file of this chapter template \texttt{abstract})
and include them with the source files of your manuscript. Use the plain \texttt{abstract} command if the abstract is also to appear in the printed version of the book.}

\abstract{We review the superfield formulation of $\mathcal{N}=2$ higher-spin supergravity theory in harmonic superspace.
The analysis of both the hypermultiplet higher-spin supersymmetries and conformal
supersymmetries is performed. The analytic superspace gauging of these symmetries gives rise to a set of unconstrained analytical prepotentials describing $\mathcal{N}=2$ higher-spin off-shell supermultiplets. This procedure
naturally yields cubic interaction vertices of $\mathcal{N}=2$ higher spins with the hypermultiplet. Based on these results, the consistent interaction of an infinite tower of $\mathcal{N}=2$ superconformal higher spins
with hypermultiplet is presented. Proceeding from this model, a method to construct a consistent interacting theory of $\mathcal{N}=2$
higher-spin supergravity by making use of the conformal compensators is proposed.}

\section{Introduction}
\label{sec:1}
The basic scientific interests
of Professor A.A. Starobinsky throughout his research career were compelled to the problems associated with
the modern development of the gravity theory and cosmology, the areas to which he made a fundamental contribution.
He also considered it useful studying the features of cosmological evolution by taking into account possible supersymmetry and/or consequences arising from generalized gravitational theories (see e.g., \cite{St1, St2, St3, St4} and reference therein).

One of the contemporary generalizations and extensions of the theory
of gravity is  related to the synthesis of the concepts of supersymmetry and higher spin fields. It is expected that such a synthesis can form the basis for
a consistent quantum gravity and the description of the Universe at
Planck scales, where all local and global symmetries are unbroken. This paper is devoted to a brief review of one of the
trends in the supersymmetric higher spin field theory.

The promising modification of the gravity
theory  is the higher-spin field theory (see reviews \cite{Vasiliev:1990en, Vasiliev:1992av, Vasiliev:1999ba}),
which contains, in addition to the conventional fields with spins
$\leq 2$, also fields with spins $> 2$. The presence of
at least one spin bigger than two in a consistent higher-spin theory
immediately leads to the need to include all the higher spins, requires
an infinite dimensional algebra of gauge symmetries and also entails
modification of Riemannian geometry, which in itself proves to be  not appropriate for
the higher spin field theory (see, e.g.,
\cite{Vasiliev:2014vwa, Tomasiello:2024jyu} and
references therein). Despite the significant progress in this theory (see e.g. \cite{Bekaert:2004qos,
Bekaert:2010hw, Bekaert:2022poo}), there still remain many open
problems. One of the most important problems is to explore the
structure of the higher-spin field interactions (see, e.g.,
recent papers \cite{Tatarenko:2024csa, Didenko:2024zpd} and
references therein).

A possible starting point of such studies is the Vasiliev
nonlinear equations for the higher-spin fields, the Lagrangian
formulation of which is still unknown.  Nevertheless, it can be
expected that such a formulation exists. Then one of the
ways to get insight into the form of admissible Lagrangians is to further limit
the set of feasible field structures. We will consider the limitations imposed by
supersymmetry, which requires the introduction of superspace for
its manifest realization (see, e.g., \cite{Wess:1992cp,
BK-book}). The superspace formulations automatically impose
restrictions on the structure of such theories and thereby
significantly limit the possible interactions and forms of
the Lagrangian description.

The most severe supersymmetric restrictions are naturally
due to extended supersymmetries realized in extended superspaces.
To date, the best formulation of theories with extended
supersymmetry is achieved within the framework of
$\mathcal{N}=2$ harmonic superspace,
discovered in 1984 \cite{Galperin:1984av, 18}. Subsequent developments have demonstrated the
fundamental role of the concept of Grassmann analyticity and the
harmonic condition of zero curvature in $\mathcal{N}=2$ theories.
These concepts arose due to the introduction of auxiliary harmonic
coordinates and have no analogues in $\mathcal{N}=0$ and
$\mathcal{N}=1$ theories. It was natural to expect that these
fundamental concepts play the fundamental role in
$\mathcal{N}=2$ higher-spin supergravity theories as well.

\smallskip

In this paper we give an overview of the results obtained in
the harmonic superspace formulation of $\mathcal{N}=2$ higher spins theories\cite{Buchbinder:2021ite,
Buchbinder:2022kzl, Buchbinder:2022vra, Buchbinder:2024pjm}. Based upon this, we make a few assumption concerning
the structure of the Lagrangians describing $\mathcal{N}=2$ higher-spin supergravity.

\smallskip

The structure of the paper is as follows. Section \ref{eq: sec harm
superspace} contains the introductory information about harmonic
superspace. In section \ref{sec:2} we describe the structure of
global higher-spin symmetries of the free hypermultiplet action.
Section \ref{sec: N=2 sugra} is devoted to the discussion of
the analytic superspace gauging of $\mathcal{N}=2$ supersymmetry and
$\mathcal{N}=2$ conformal supersymmetry, which leads to analytic
prepotentials of $\mathcal{N}=2$ Einstein supergravity and
$\mathcal{N}=2$ conformal supergravity. We compare the
supermultiplets obtained, discuss the relevant supersymmetric actions and the method of
conformal compensators, which allows  to build Einstein
supergravities from the conformal one. In section \ref{eq: sec HS} we show
that the analytical prepotentials of supergravity naturally
generalize to higher spins.  In section \ref{sec: consistent} we
discuss a consistent model of the hypermultiplet interacting with an infinite
tower of $\mathcal{N}=2$ superconformal higher-spins and outline a number of
the related open problems and possible applications. The concluding
section \ref{sec: 7} involves discussions of the possible structure
of $\mathcal{N}=2$ higher-spin supergravity, as well as poses some
open problems related to the construction of such a theory. For
convenience of the reader, in the appendix we collect  the notation
used throughout the paper.

We apologize for inevitable omissions in the reference list caused by the limited size of the Contribution.
More complete list can be found in our original papers on the subject.

\section{Harmonic superspace}
\label{eq: sec harm superspace}

The supersymmetric field theories, in addition to the standard Poincar\`e invariance, respect a supersymmetry, the invariance under transformations with fermionic parameters
mixing bosonic and fermionic fields. To construct theories with the manifestly realized supersymmetry, the Minkowski space should be extended by anticommuting fermionic coordinates
$\theta^\alpha_i$, $\bar{\theta}_{\dot{\alpha}}^i$ \cite{Wess:1992cp,BK-book}. This is  $\mathcal{N}$-extended Minkowski superspace
\begin{equation}
    \mathbb{R}^{4|4\mathcal{N}} = \{ x^{\alpha\dot{\alpha}}, \theta^\alpha_i, \bar{\theta}^i_{\dot{\alpha}}\}, \qquad i =1, \dots \mathcal{N}.
\end{equation}
In superspace, supersymmetry is geometrically implemented  by the coordinate transformations
\begin{equation}\label{eq: susy R44}
    \delta_\epsilon x^{\alpha\dot{\alpha}} =
    2i \left( \epsilon^{\alpha i} \bar{\theta}^{\dot{\alpha}}_i
    -
    \theta^{\alpha i} \bar{\epsilon}^{\dot{\alpha}}_i
    \right),
    \qquad
    \delta \theta^\alpha_i = \epsilon^\alpha_i,
    \qquad
    \delta \bar{\theta}^i_{\dot{\alpha}} = \bar{\epsilon}^i_{\dot{\alpha}}.
\end{equation}
The analogs of relativistic fields in superspace are superfields $\Phi(x, \theta, \bar{\theta})$. When being expanded over Grassmann coordinates $\theta$ and $\bar{\theta}$,
these amount to finite sets of ordinary fields.
The action principle, gauge transformations, symmetries - everything can be described in terms of superfields \cite{Wess:1992cp, BK-book}.
The superspace formulations of supersymmetric theories have the  structure different from that of  non-supersymmetric theories. Moreover, it turns out that
the superspace $\mathbb{R}^{4|4\mathcal{N}}$ suits only for describing theories with the minimal $\mathcal{N}=1$ supersymmetry.
For the full-fledged formulations of theories with extended $\mathcal{N}\geq 2$ supersymmetry,  some modifications of the underlying superspace are required.

The most developed version of the superspaces for extended supersymmetry is $\mathcal{N}=2$ harmonic superspace \cite{Galperin:1984av, 18}.
The harmonic superspace involves additional bosonic coordinates, harmonics $u^\pm_i$ which parametrize the auxiliary sphere $S^2 \sim SU(2)/U(1)$. The harmonics satisfy the relation $u^{+i} u^-_i = 1$
and allow, like vielbeins in the general relativity, to convert $SU(2)$ indices into the $U(1)$ ones. In the harmonic superspace
\begin{equation}
    \mathbb{HR}^{4+2|8} = \{x^{\alpha\dot{\alpha}}, \theta^{\alpha i}, \bar{\theta}^{i}_{\dot{\alpha}}, u^\pm_i \},
    \qquad
    i = 1,2\,,
\end{equation}
superfields $\Phi^{(n)} (x, \theta, \bar{\theta}, u)$ with a fixed $U(1)$ charge $``n''$ are considered. Unlike ordinary superfields, the harmonic superfields contain infinitely many component
fields which appear from the harmonic expansions.

The introduction of harmonics makes it possible to distinguish a new superspace closed under the supersymmetry -- the \textit{analytic superspace} containing half of the original fermionic coordinates:
\begin{equation}\label{eq: analyt SS}
    \mathbb{HA}^{4+2|4} =\{\zeta\} = \{ x_A^{\alpha\dot{\alpha}}, \theta_A^{+\alpha}, \bar{\theta}_A^{+\dot{\alpha}}, u^\pm_i \}.
\end{equation}
Here we have introduced the analytic basis coordinates:
\begin{equation}
    x_A^{\alpha\dot{\alpha}} = x^{\alpha\dot{\alpha}}- 4i \theta^{\alpha (i} \bar{\theta}^{\dot{\alpha}j)} u^+_i u^-_j,
    \quad
    \theta_A^{\pm\alpha} =\theta^{\alpha i} u^\pm_i ,
    \quad
    \bar{\theta}_A^{\pm\dot{\alpha}} = \bar{\theta}^{\dot{\alpha} i} u^\pm_i.
\end{equation}
Under $\mathcal{N}=2$ supersymmetry \eqref{eq: susy R44} the analytic coordinates transform as
\begin{equation}\label{eq: rig susy}
    \delta_\epsilon x_A^{\alpha\dot{\alpha}} = -4i \left(\epsilon^{-\alpha} \bar{\theta}^{+\dot{\alpha}} + \theta^{+\alpha} \bar{\epsilon}^{-\dot{\alpha}} \right),
    \;
    \delta_\epsilon \theta_A^{\pm\alpha} = \epsilon^{\pm\alpha},
    \;
    \delta_\epsilon \bar{\theta}_A^{\pm\dot{\alpha}}
    = \bar{\epsilon}^{\pm\dot{\alpha}},
    \;
    \delta_\epsilon u^\pm_i = 0,
\end{equation}
so the analytic superspace \eqref{eq: analyt SS} is invariant under $\mathcal{N}=2$ supersymmetry. Also, the analytic superspace is real with respect
to the tilde conjugation (eq. \eqref{eq: tilde conj} in what follows), which generalizes the usual complex conjugation.

To describe a massive hypermultiplet (with the mass equal to central charge) it is also useful to introduce the auxiliary central charge coordinate $x^5$
with the $\mathcal{N}=2$ supersymmetry transformations law $\delta_\epsilon x_5 = - i (\epsilon^{\alpha i} \theta_{\alpha i} - \bar{\theta}^i_{\dot{\alpha}} \bar{\theta}^{\dot{\alpha}}_i)$ \cite{Galperin:1984av, 18}.
This is an example of the Scherk-Schwartz dimension reduction method \cite{Scherk:1979zr}.  In the analytic basis we
introduce $x^5_A = x^5 + i (\theta^{+\alpha}\theta^-_{\alpha} - \bar{\theta}^+_{\dot{\alpha}} \bar{\theta}^{-\dot{\alpha}} )$ with the analyticity-preserving transformation law under supersymmetry:
\begin{equation}\label{eq: x5 susy}
    \delta_\epsilon x_A^5 = 2i (\epsilon^{-\alpha}\theta^{+}_\alpha - \bar{\epsilon}^-_{\dot{\alpha}} \bar{\theta}^{+\dot{\alpha}}).
\end{equation}
Hereafter, unless specified, we consider $x^5$-independent superfields,  work in the analytic basis and  omit the subscript $A$.

The presence of harmonics allows one to introduce
\textit{harmonic derivatives} $\partial^{\pm\pm} = u^{\pm i}\frac{\partial}{\partial u^{\mp i}}, \partial^0 = u^{+i}\frac{\partial}{\partial u^{+i}} - u^{-i}\frac{\partial}{\partial u^{-i}}$,
which are consistent with the harmonic defining relation $u^{+i}u^-_i = 1$. In the analytic basis, the harmonic derivatives take the form:
\begin{subequations}
\begin{equation}
    \mathcal{D}^{\pm\pm} = \partial^{\pm\pm} - 4i \theta^{\pm\alpha} \bar{\theta}^{\pm\dot{\alpha}} \partial_{\alpha\dot{\alpha}} + \theta^{\pm\hat{\alpha}} \partial^{\pm}_{\hat{\alpha}}
    +
    i \left[(\theta^+)^2 - (\bar{\theta}^+)^2\right] \partial_5,
\end{equation}
\begin{equation}
    \mathcal{D}^0 = \partial^0 + \theta^{+\hat{\alpha}} \partial^-_{\hat{\alpha}} - \theta^{-\hat{\alpha}} \partial^+_{\hat{\alpha}}.
\end{equation}
\end{subequations}
The harmonic derivatives are invariant under rigid $\mathcal{N}=2$ supersymmetry \eqref{eq: rig susy}, \eqref{eq: x5 susy}.

The analytic superspace and harmonic derivatives play the key role in the superspace formulation of $\mathcal{N}=2$ theories \cite{Galperin:1984av, 18}.
In the next sections, we will demonstrate that the analyticity concept in conjunction with the harmonic derivatives severely constrain the structure of $\mathcal{N}=2$ supergravity,
as well as of $\mathcal{N}=2$ higher-spin theories. We will pay a special attention to the ways how it is accomplished.

\section{Hypermultiplet and higher-spin symmetries in $\mathcal{N}=2$ harmonic superspace}
\label{sec:2}

The fundamental $\mathcal{N}=2$ matter supermultiplet, the hypermultiplet, is described in harmonic superspace  by an unconstrained analytic superfield $q^+(\zeta)$
with the free action \begin{equation}\label{eq: free hyper}
    S_{hyp} = - \int d\zeta^{(-4)} \tilde{q}^+ \mathcal{D}^{++} q^+
    =
    - \frac{1}{2} \int d\zeta^{(-4)} q^{+a} \mathcal{D}^{++} q^+_a.
\end{equation}
Here we used  the notation $q^+_a = (q^+, - \tilde{q}^+)$, $q^{+a} = \epsilon^{ab}q^+_b$. In its second form, the action \eqref{eq: free hyper} is manifestly invariant
under $SU(2)_{PG}$ group that acts on the doublet indices $a$.

The mass of the hypermultiplet can be introduced by allowing for a dependence on the central charge coordinate $x^5$
\begin{equation}
    q^+(\zeta)\quad \to \quad q^+(\zeta, x^5) = e^{imx^5}   q^+(\zeta).
\end{equation}
The hypermultiplet mass is equal to the central charge, otherwise a massive hypermultiplet cannot be defined.
Note that, due to the form of $x^5$-dependence, the Lagrangian density in \eqref{eq: free hyper} is $x^5$-independent.

In accordance with the general features of harmonic superfields, the hypermultiplet superfield contains an infinite number of component fields in its $\theta$ and $u$ expansions.
However, almost all the component fields are auxiliary and become zero on the free equations of motion $\mathcal{D}^{++}q^+ = 0$, leaving in $q^+ $ only the physical ones:
\begin{equation}
 q^+(\zeta) = f^i u^+_i + \theta^{+\alpha} \psi_\alpha + \bar{\theta}_{\dot{\alpha}} \bar{\chi}^{\dot{\alpha}} + m \left[ (\theta^+)^2 - (\bar{\theta}^+)^2 \right] f^i u^-_i
 +
 4i \theta^{+\alpha}\bar{\theta}^{+\dot{\alpha}} \partial_{\alpha\dot{\alpha}} f^i u^-_i.
\end{equation}

\medskip

One of the central topics in the theory of higher spins is the higher-spin (HS) (super)symmetries. In a certain sense, the symmetry of higher spins
is the maximally possible relativistic symmetry and the low-spin symmetries can be considered as low-energy symmetries inherent to the spontaneously broken phase.
It turns out that the higher-spin symmetries of similar kind appear in the simplest relativistic equations.
The higher-spin symmetries of the free massless scalar field were discovered by Shaynkman and Vasiliev in $3d$ \cite{Shaynkman:2001ip} and by Eastwood for any $d$ \cite{Eastwood:2002su}
(see also \cite{Vasiliev:2014vwa}). Therefore, it is quite natural to pose the problem of investigating the higher-spin (super)symmetries of the free hypermultiplet action \eqref{eq: free hyper},
which is just $\mathcal{N}=2$ analogue of the free Klein-Gordon action.

\smallskip

Further in this section we will present some details related to the higher-spin $\mathcal{N}=2$ supersymmetries of  the massive hypermultiplet and to $\mathcal{N}=2$ conformal supersymmetries
of the massless hypermultiplet. Then, in sections \ref{sec: N=2 sugra} and \ref{eq: sec HS}, we will demonstrate that the analytic gauging of such symmetries
naturally leads to the $\mathcal{N}=2$ higher spin gauge superfields.

\medskip

\subsubruninhead{Rigid $\mathcal{N}=2$ Poincar\'e supersymmetry}

\medskip

The free hypermultiplet action \eqref{eq: free hyper} is manifestly invariant under $\mathcal{N}=2$ supersymmetry \eqref{eq: rig susy}, \eqref{eq: x5 susy}:
\begin{equation}
    \delta^\star_\epsilon q^+ := q^{\prime +}(\zeta^\prime) - q^{+} (\zeta) = 0.
\end{equation}
For the higher-spin generalization, it is useful to consider the active form of this supersymmetry variation
\begin{equation}\label{eq: hyper susy}
    \delta_\epsilon q^+ := q^{\prime +} (\zeta) - q^+(\zeta) = - \hat{\Lambda}_{susy} q^{+}.
\end{equation}
Here we introduced the first-order differential operator
\begin{equation}\label{eq: susy}
    \begin{split}
    \hat{\Lambda}_{susy} = &\Big\{ a^{\alpha\dot{\alpha}}
    +
     l^{(\alpha}_{\;\;\beta)} x^{\beta\dot{\alpha}} + \bar{l}^{(\dot{\alpha}}_{\;\;\dot{\beta})} x^{\alpha\dot{\beta}}
     -4i \left(\epsilon^{\alpha i}u^-_i \bar{\theta}^{+\dot{\alpha}} + \theta^{+\alpha} \bar{\epsilon}^{\dot{\alpha}i}u^-_i \right) \Big\} \partial_{\alpha\dot{\alpha}}
    \\&+ \left\{ \epsilon^{\alpha i} u^+_i + l^{(\alpha}_{\;\;\beta)} \theta^{+\beta}\right\} \partial^-_{\alpha}
    +  \left\{
    \bar{\epsilon}^{\dot{\alpha}i}u^+_i
    +
     \bar{l}^{(\dot{\alpha}}_{\;\;\dot{\beta})} \bar{\theta}^{+\dot{\beta}}\right\}
    \partial^-_{\dot{\alpha}}
        \\&+ \left\{ \epsilon^{\alpha i} u^-_i + l^{(\alpha}_{\;\;\beta)} \theta^{-\beta}\right\} \partial^+_{\alpha}
    +  \left\{
    \bar{\epsilon}^{\dot{\alpha}i}u^-_i
    +
    \bar{l}^{(\dot{\alpha}}_{\;\;\dot{\beta})} \bar{\theta}^{-\dot{\beta}}\right\}
    \partial^+_{\dot{\alpha}}
    \\
    & + \left\{c + 2i (\epsilon^{\alpha i} u^-_i\theta^{+}_\alpha - \bar{\epsilon}^i_{\dot{\alpha}} u^-_i \bar{\theta}^{+\dot{\alpha}}) \right\}\partial_5.
    \end{split}
\end{equation}
The parameters $a^{\alpha\dot{\alpha}}$ correspond to rigid translations, $l_{(\alpha\beta)}$ and $l_{(\dot{\alpha}\dot{\beta})}$ to Lorenz transformations,
$\epsilon^{\alpha\pm} = \epsilon^i u^\pm_i, \bar{\epsilon}^{\dot{\alpha}\pm} = \bar{\epsilon}^{\dot{\alpha}i}u^\pm_i$ are the supersymmetry parameters,
the parameter $c$ corresponds to $U(1) \subset SU(2)_{PG}$ symmetry. One can easily check that Lie brackets of these transformations form $\mathcal{N}=2$ super-Poincar\'e algebra.

The operator $\hat{\Lambda}_{susy}$ is the unique first-order superspace differential operator that satisfies the relation
\begin{equation}\label{eq: rigid susy}
    [\mathcal{D}^{++},  \hat{\Lambda}_{susy}] = 0.
\end{equation}
This relation amounts to the invariance of the harmonic derivative under $\mathcal{N}=2$ supersymmetry and ensures the invariance of the massive hypermultiplet action \eqref{eq: free hyper}
under $\mathcal{N}=2$ supersymmetry transformations in the active form \eqref{eq: hyper susy}.

\subsubruninhead{Rigid $\mathcal{N}=2$ superconformal symmetry}

\medskip

In the massless case, one can expect that the symmetry group is larger and should include conformal invariance.
Accordingly, one may consider a more general condition for the hypermultiplet rigid symmetries:
\begin{equation}\label{eq: N=2 sc}
    [\mathcal{D}^{++}, \hat{\Lambda}_{sc}] = \lambda^{++}_{sc} \mathcal{D}^0.
\end{equation}
Since $q^{+a}\mathcal{D}^0 q^+_a = 0$, this condition implies the invariance of the free hypermultiplet action. The transformations of $q^+_a$ which lead to the variation
\begin{equation}
    \delta_{sc} S_{hyp} = \frac{1}{2}  \int d\zeta^{(-4)} q^{+a}    [\mathcal{D}^{++}, \hat{\Lambda}_{sc}] q^+_a = 0\,,
\end{equation}
are found to have the form:
\begin{equation}\label{eq: sc transf}
    \delta_{sc} q^{+}_a = - \hat{\Lambda}_{sc} q^{+}_a - \frac{1}{2} \Omega_{sc} q^{+}_a,
\end{equation}
where we have introduced the differential operator
\begin{equation}\label{eq: Lambda sc}
    \hat{\Lambda}_{sc} = \lambda_{sc}^{\alpha\dot{\alpha}}\partial_{\alpha\dot{\alpha}}
    +
    \lambda^{+\alpha}_{sc} \partial^-_{\alpha}
    +
    \lambda^{+\dot{\alpha}}_{sc} \partial^-_{\dot{\alpha}}
    +
    \lambda^{-\alpha}_{sc} \partial^+_{\alpha}
    +
    \lambda^{-\dot{\alpha}}_{sc} \partial^+_{\dot{\alpha}}
    +
    \lambda^{++} \partial^{--},
\end{equation}
and the weight factor
\begin{equation}
    \Omega_{sc} =
    \partial_{\alpha\dot{\alpha}} \lambda_{sc}^{\alpha\dot{\alpha}}
    -
    \partial^-_{\alpha}
    \lambda^{+\alpha}_{sc}
    -
    \partial^-_{\dot{\alpha}}
    \lambda^{+\dot{\alpha}}_{sc}
    +
    \partial^{--} \lambda^{++}.
\end{equation}
The operator $\hat{\Lambda}_{sc}$,  in contrast to $\hat{\Lambda}_{susy}$, contains the harmonic derivative $\partial^{--}$
but lacks the  central charge derivative $\partial_5$. It is straightforward to verify that a non-trivial  $\partial_5$ term that corresponds to the hypermultiplet mass
does not allow solutions with non-vanishing $\lambda^{++}$ parameter \cite{Buchbinder:2024pjm}.

As a solution to the condition \eqref{eq: N=2 sc} we obtain the parameters $\lambda^M_{sc} $ (to which one should add the rigid super-Poincar\'e parameters
from \eqref{eq: susy} without the last line, corresponding to the case of trivial central charge,  $\partial_5q^+= 0$):
\begin{equation}\label{eq:superconformal symmetry}
\begin{split}
%   \begin{cases}
        \lambda_{sc}^{\alpha\dot{\alpha}}
        =&
        x^{\dot{\alpha}\rho} k_{\rho\dot{\rho}} x^{\dot{\rho}\alpha}
    %   \\&
        +
        d x^{\alpha\dot{\alpha}}
        -
        4i \theta^{+\alpha} \bar{\theta}^{+\dot{\alpha}} \lambda^{(ij)}u^-_i u^-_j
        -
        4i \left(x^{\alpha\dot{\rho}}\bar{\eta}_{\dot{\rho}}^i \bar{\theta}^{+\dot{\alpha}}
        +
        \theta^{+\alpha} \eta^i_{\rho} x^{\rho\dot{\alpha}}
        \right) u^-_i,
        \\
        \lambda^{+\alpha}_{sc}
        =&
        \frac{1}{2}
        \theta^{+\alpha} (d + i r)
        +
        x^{\alpha\dot{\beta}} k_{\beta\dot{\beta}} \theta^{+\beta}
        +
        x^{\alpha\dot{\alpha}}  \eta^i_{\dot{\alpha}} u^+_i
    %   \\&
        +
        \theta^{+\alpha}
        \left( \lambda^{(ij)}u^+_i u^-_j
        +
        4i \theta^{+\rho} \eta^i_{\rho} u^-_i
        \right),
        \\
        \bar{\lambda}^{+\dot{\alpha}}_{sc}
        =&
        \frac{1}{2}
        \bar{\theta}^{+\dot{\alpha}}  (d - i r)
        +
        x^{\dot{\alpha}\beta} k_{\beta\dot{\beta}} \bar{\theta}^{+\dot{\beta}}
        +
        x^{\alpha\dot{\alpha}}  \eta^i_{\alpha} u^+_i
    %   \\&
        +
        \bar{\theta}^{+\dot{\alpha}}
        \left( \lambda^{(ij)}u^+_i u^-_j
        -
        4i \bar{\theta}^{+\dot{\rho}} \bar{\eta}^i_{\dot{\rho}} u^-_i
        \right),
        \\
        \lambda^{-\alpha}_{sc} =\;&
        \frac{1}{2} \theta^{-\alpha} (d+ir)
        +
        x^{\alpha\dot{\beta}}
        k_{\beta\dot{\beta}}   \theta^{-\beta}
        -
        2i (\theta^-)^2 \bar{\theta}^+_{\dot{\beta}} k^{\dot{\beta}\alpha}
    + \left( x^{\alpha\dot{\alpha}} + 4i \theta^{-\alpha}\bar{\theta}^{+\dot{\alpha}} \right) \bar{\eta}^i_{\dot{\alpha}} u^-_i
            \\& +
        4i \eta^i_{\beta} \theta^{-\beta} \left( \theta^{-\alpha} u^+_i - \theta^{+\alpha} u^-_i \right)
        +
        \lambda^{ij} u^-_i \left( u^-_j \theta^{+\alpha} - u^+_j \theta^{-\alpha} \right),
        \\
        \lambda^{-\dot{\alpha}}_{sc}
        =\;&
        \frac{1}{2} \bar{\theta}^{-\dot{\alpha}}(d-ir)
        +
        x^{\dot{\alpha}\beta} k_{\beta\dot{\beta}} \bar{\theta}^{-\dot{\beta}}
        -
        2i (\bar{\theta}^-)^2 \theta^+_{\beta} k^{\beta\dot{\alpha}}
        +
        \left( x^{\alpha\dot{\alpha}} + 4i \theta^{+\alpha} \bar{\theta}^{-\dot{\alpha}} \right) \eta^i_{\alpha} u^-_i
        \\&
        -
        4i \bar{\eta}^i_{\dot{\beta}} \bar{\theta}^{-\dot{\beta}} \left( \bar{\theta}^{-\dot{\alpha}} u^+_i - \bar{\theta}^{+\dot{\alpha}} u^-_i \right)
        +
        \lambda^{ij} u^-_i \left( u^-_j \bar{\theta}^{+\dot{\alpha}} - u^+_j \bar{\theta}^{-\dot{\alpha}} \right),
        \\
        \lambda^{++}_{sc} =&
        \lambda^{ij} u^+_i u^+_j
        +
        4i \theta^{+\alpha} \bar{\theta}^{+\dot{\alpha}} k_{\alpha\dot{\alpha}}
        +
        4i \left( \theta^{+\alpha} \eta^i_{\alpha} + \bar{\eta}^i_{\dot{\alpha}} \bar{\theta}^{+\dot{\alpha}}  \right) u^+_i.
%   \end{cases}
\end{split}
\end{equation}
Here the parameters $k_{\alpha\dot{\alpha}}$  $d$ , $r$, $\lambda^{(ij)}$ and $\eta^i_\alpha, \bar{\eta}^i_{\dot{\alpha}}$ are associated with the special conformal transformations, dilatations, $U(1)_R$ symmetry,
$SU(2)_R$ symmetry and the conformal supersymmetry, respectively. The transformations  \eqref{eq: susy} (without the last line)   and \eqref{eq:superconformal symmetry}
form $\mathcal{N}=2$ superconformal algebra $su(2,2|2)$.

\medskip

\subsubruninhead{Rigid $\mathcal{N}=2$ higher-spin supersymmetry}

\medskip

The active form of $\mathcal{N}=2$ supersymmetry transformations \eqref{eq: hyper susy} of the massive hypermultiplet  is convenient, when seeking for generalizations to higher spins.
We introduce the differential operator involving the higher-order  space-time derivatives:
\begin{equation}\label{eq: spin s susy}
    \hat{\Lambda}_{susy}^{(s)} = \lambda_{susy}^{\alpha(s-2)\dot{\alpha}(s-2)M} \partial_M \partial^{s-2}_{\alpha(s-2)\dot{\alpha}(s-2)}.
\end{equation}
Here $M=\{\alpha\dot{\alpha}, \alpha+, \dot{\alpha}+, \alpha-, \dot{\alpha}-, 5 \}$ and we assume that all the Lorentz indices of the same type  are symmetrized.

As in the case of $\mathcal{N}=2$ supersymmetry \eqref{eq: rigid susy}, we impose the condition
\begin{equation}
    [\mathcal{D}^{++}, \hat{\Lambda}_{susy}^{(s)}] =  0.
\end{equation}
As its general solution, we obtain:
\begin{equation}
    \begin{split}
        \lambda_{susy}^{\alpha(s-1)\dot{\alpha}(s-1)} &= a^{\alpha(s-1)\dot{\alpha}(s-1)}
    %   \\&\;\;\;
    +
        l^{(\dot{\alpha}(s-2)(\alpha(s-1)}_{\qquad\qquad\quad\;\;\beta)} x^{\beta\dot{\alpha})}
        +
         \bar{l}^{(\alpha(s-2)(\dot{\alpha}(s-1)}_{\qquad \qquad\quad\;\;\dot{\beta})} x^{\alpha)\dot{\beta}}
        \\&
        \quad- 4i \left(
        \epsilon^{\alpha(s-1)\dot{\alpha}(s-2)i} u^-_i \bar{\theta}^{+\dot{\alpha}}
        +
        \theta^{+(\alpha} \bar{\epsilon}^{\alpha(s-2))\dot{\alpha}(s-1)i} u^-_i
        \right),
        \\
        \lambda_{susy}^{\pm\alpha(s-1)\dot{\alpha}(s-2)}
        &=
        \epsilon^{\alpha(s-1)\dot{\alpha}(s-2) i} u_i^\pm
        +
        l^{\dot{\alpha}(s-2)(\alpha(s-1)}_{\qquad\qquad\;\;\;\;\beta)} \theta^{\pm \beta},
            \\
        \lambda_{susy}^{\pm\alpha(s-2)\dot{\alpha}(s-1)}
        &=
        \epsilon^{\alpha(s-2)\dot{\alpha}(s-1) i} u_i^\pm
        +
        \bar{l}^{\alpha(s-2)(\dot{\alpha}(s-1)}_{\qquad\qquad\;\;\;\;\dot{\beta})} \bar{\theta}^{\pm \dot{\beta}},
        \\
        \lambda_{susy}^{\alpha(s-2)\dot{\alpha}(s-2)5} & =
        c^{\alpha(s-2)\dot{\alpha}(s-2)}
        +
        2i \epsilon^{\alpha(s-2)\dot{\alpha}(s-2)\hat{\alpha}i}u^-_i \theta^+_{\hat{\alpha}}.
    \end{split}
\end{equation}
For $s=2$ these parameters are reduced just to the rigid $\mathcal{N}=2$ supersymmetry \eqref{eq: hyper susy}.
So the relevant transformations provide the natural  higher-spin generalization of $\mathcal{N}=2$ supersymmetry.
The parameters $a^{\alpha(s-1)\dot{\alpha}(s-1)}$ and $c^{\alpha(s-2)\dot{\alpha}(s-2)}$ are the higher-spin generalizations
of translation and $U(1)$ parameters, $\epsilon^{\alpha(s-2)\dot{\alpha}(s-2)i}$ are the parameters of $\mathcal{N}=2$ higher-spin supersymmetry.

It is interesting to consider the algebraic structure of these transformations:
\begin{equation}
    [\hat{\Lambda}^{(s_1)}_{susy}, \hat{\Lambda}^{(s_2)}_{susy}] \sim \hat{\Lambda}_{susy}^{(s_1+s_2-2)}.
\end{equation}
For the particular case of $s_1=s_2=2$ we reproduce the standard closed $\mathcal{N}=2$ Super-Poincar\'e algebra. However, if there is at least one spin $s>2$,
generalized $\mathcal{N}=2$ supersymmetry transformations of arbitrary spin appear.

\smallskip

The operators defined above generate the higher-spin supersymmetries of the hypermultiplet. Actually, the hypermultiplet transformations that lead to the vanishing variation
\begin{equation} \delta_{susy}^{(s)} S_{hyp} = \frac{1}{2}  \int d\zeta^{(-4)} q^{+a}   [\mathcal{D}^{++}, \hat{\Lambda}_{susy}^{(s)}] q^+_a = 0
\end{equation}
are given by
\begin{equation} \label{eq: spin s h susy}
    \delta^{(s)}_{susy} q^+_a = - \hat{\Lambda}_{susy}^{(s)} (J)^{P(s)} q^+_a,
\end{equation}
where we have introduced the $U(1)$ generator $J$ and the sign operator $P(s)$:
\begin{equation}
    J q^+ :=  i q^{+}, \qquad J  \tilde{q}^+ :=- i \tilde{q}^+,
    \qquad
    P(s) :=
    \begin{sqcases}
        0 \quad \text{for even} \; s ;
        \\
        1 \quad \text{for odd}\; s .
    \end{sqcases}
\end{equation}
The term $(J)^{P(s)}$ in \eqref{eq: spin s h susy} distinguishes between the charges of the hypermultiplet $q^+$ and its tilde-conjugate $\tilde{q}^+$,
when considering  the odd higher spin transformations.  This is the higher-spin reflection of the well-known fact that
the minimal interaction of an electromagnetic field is possible only for a complex scalar field, the field and its complex conjugate
having opposite electric charges.

\medskip

\subsubruninhead{Rigid $\mathcal{N}=2$ higher-spin superconformal symmetry}

\medskip

Like in the case of $\mathcal{N}=2$ superconformal symmetry, the number of higher-spin symmetries for a massless hypermultiplet increases.
As in $\mathcal{N}=2$ superconformal transformations, here we also add harmonic derivative $\partial^{--}$ and drop terms with the central charge operator $\partial_5$.
 The general ansatz for transformations has the form
\begin{equation}
    \hat{\Lambda}^{(s)}_{sc} = \lambda^{M_1\dots M_{s-1}} \partial_{M_{s-1}} \dots \partial_{M_1} + \text{lower derivative contributions}.
\end{equation}
Instead of performing a complete analysis of these transformations and the relevant symmetry conditions, we will confine our attention
to  a non-trivial example of the  higher-spin dilatation transformations:
\begin{equation}
    \begin{split}
    \delta_{dil} q^+_a  = &- d^{\alpha(s-2)\dot{\alpha}(s-2)} \left[x^{\alpha\dot{\alpha}} \partial_{\alpha\dot{\alpha}} + \frac{1}{2} \theta^{+\hat{\alpha}} \partial^+_{\hat{\alpha}}  \right]  \partial^{s-2}_{\alpha(s-2)\dot{\alpha}(s-2)}  (J)^{P(s)}q^+_a
    \\
    &\;+ d^{\alpha(s-2)\dot{\alpha}(s-2)} \partial^{s-2}_{\alpha(s-2)\dot{\alpha}(s-2)} (J)^{P(s)}  q^+_a.
    \end{split}
\end{equation}
The special conformal transformations and the conformal supersymmetry are generalized in a similar way.

\medskip

Now, having described the structure of global higher-spin $\mathcal{N}=2$ supersymmetries, we can move on
to their analytic gauging and the definition of the corresponding gauge $\mathcal{N}=2$ superfields.

\section{$\mathcal{N}=2$ supergravity}
\label{sec: N=2 sugra}

Before proceeding to  $\mathcal{N}=2$ higher spin supergravity, we consider an instructive
example of $\mathcal{N}=2$ supergravity \cite{18, Galperin:1987em, Galperin:1987ek,  Ivanov:2022vwc}. This is a simple example
of the consistent gauge theory with finite set of fields, which is a useful prototype of the higher-spin supergravities.
The harmonic formulation of $\mathcal{N}=2$ supergravity suggests some conjectures concerning the structure of the complete   $\mathcal{N}=2$ higher-spin theory.

\medskip

\subsubruninhead{$\mathcal{N}=2$ Einstein supergravity}

\medskip

To obtain a superfield description of the $\mathcal{N}=2$ supergravity multiplet, we consider the gauged version  of rigid $\mathcal{N}=2$ supersymmetry transformations \eqref{eq: susy}.
Actually, since the gauge transformations contain local translations, the invariance of the action with respect to such transformations requires the introduction of gravitational field.
$\mathcal{N}=2$ supersymmetry further requires that the gravitational field is accompanied by other fields from $\mathcal{N}=2$ supergravity multiplet.

We demand that the gauged transformations preserve the analytical superspace \eqref{eq: analyt SS}.
So, the analytic gauging of rigid transformations takes the form:
\begin{equation}\label{eqL super diff}
    \begin{split}
    \hat{\Lambda}_{susy}
    \quad
    \to
    \quad
    \hat{\Lambda}^{loc}_{susy} = &\lambda^{\alpha\dot{\alpha}}(\zeta) \partial_{\alpha\dot{\alpha}}
    +
    \lambda^{\hat{\alpha}+} (\zeta) \partial^-_{\hat{\alpha}}
    +
    \lambda^{\hat{\alpha}-}(\zeta, \theta^-) \partial^+_{\hat{\alpha}}
    +
    \lambda^5 (\zeta) \partial_5.
    \end{split}
\end{equation}
These transformations form a closed algebra, $[\hat{\Lambda}^{loc}_{susy},  \hat{\Lambda}^{loc}_{susy} ] \sim \hat{\Lambda}^{loc}_{susy} $.

The harmonic derivative $\mathcal{D}^{++}$ is not invariant with respect to such transformations, $[ \mathcal{D}^{++},   \hat{\Lambda}^{loc}_{susy}]\neq 0$ (compare with eq. \eqref{eq: rigid susy}).
To restore the invariance, we covariantize the harmonic derivative as
\begin{equation}
    \mathcal{D}^{++} \quad \to \quad \mathfrak{D}^{++} = \mathcal{D}^{++} + \kappa_2 \hat{\mathcal{H}}^{++}_{(s=2)}, \label{CovD++}
\end{equation}
 thereby introducing the vielbeins appearing in the first-rank differential operator:
\begin{equation}\label{eq: spin 2 oper}
    \hat{\mathcal{H}}^{++}_{(s=2)} = h^{++\alpha\dot{\alpha}}(\zeta) \partial_{\alpha\dot{\alpha}}
    +
    h^{++\hat{\alpha}+} (\zeta)\partial^{-}_{\hat{\alpha}}
    +
    h^{++\hat{\alpha}-} (\zeta, \theta^-) \partial^+_{\hat{\alpha}}
    +
    h^{++5} (\zeta) \partial_5
\end{equation}
(in eq. \eqref{CovD++} $\kappa_2$ is the coupling constant, see eq. \eqref{eq: hyp Ein sg} below). The  requirement of invariance of the harmonic derivative $\mathfrak{D}^{++}$ fixes the transformation laws of the vielbeins
\begin{equation}\label{eq: spin 2 H transf}
    \begin{split}
&\delta_\lambda \mathfrak{D}^{++} =  \kappa_2   [\hat{\Lambda}^{loc}_{susy}, \mathfrak{D}^{++}]  +  \kappa_2    \delta_\lambda  \hat{\mathcal{H}}^{++}_{(s=2)} = 0
    \\
    &
        \qquad  \qquad
    \Rightarrow
    \quad \delta_\lambda  \hat{\mathcal{H}}^{++}_{(s=2)} = [\mathcal{D}^{++} + \kappa_2\hat{\mathcal{H}}_{(s=2)}, \hat{\Lambda}_{susy}^{loc}].
    \end{split}
\end{equation}

All the vielbeins are analytic, except for $h^{++\hat{\alpha}-}(\zeta, \theta^-)$; fortunately, the latter can be entirely gauged away.
 Indeed,  under the gauge transformations \eqref{eq: spin 2 H transf} the  non-analytical vielbein transforms as
\begin{equation}
    \delta_\lambda h^{++\hat{\alpha}-} = \left(\mathcal{D}^{++} +\kappa_2 \hat{\mathcal{H}}^{++}_{(s=2)} \right) \lambda^{-\hat{\alpha}}
    -
    \lambda^{+\hat{\alpha}}
    -
    \kappa_2 \hat{\Lambda}_{susy}^{loc} h^{++\hat{\alpha}-},
\end{equation}
and, since the non-analytical gauge parameter $\lambda^{-\hat{\alpha}}$ is unconstrained, one can impose the \textit{analytic gauge} $h^{++\hat{\alpha}-} = 0$.
In this gauge the covariant harmonic derivative $\mathfrak{D}^{++}$  preserves the analyticity, $[\mathcal{D}^+_{\hat{\alpha}}, \mathfrak{D}^{++}] = 0$, and
the parameter $\lambda^{-\hat{\alpha}}$ is expressed in terms of the analytic $\lambda^{+\hat{\alpha}}$.

Using the operator \eqref{eq: spin 2 oper}  one can construct  the gauge invariant coupling to the hypermultiplet:
\begin{equation}\label{eq: hyp Ein sg}
    S^{Ein}_{hyp} = -\frac{1}{2} \int d\zeta^{(-4)} q^{+a} \left( \mathcal{D}^{++} + \kappa_2 \hat{\mathcal{H}}^{++}_{(s=2)}  \right) q^+_a.
\end{equation}
This coupling is invariant under super-diffeomorphisms \eqref{eq: spin 2 H transf} accompanied by the hypermultiplet transformations
\begin{equation}
    \delta_\lambda q^{+}_a = - \kappa_2 \hat{\mathcal{U}}_{susy} q^+_a  = -\kappa_2 \left( \hat{\Lambda}^{loc}_{susy}  + \frac{1}{2} \Omega_{susy}^{loc} \right) q^+_a.
\end{equation}
Here we introduced the weight factor $\Omega_{susy}^{loc} := (-1)^{P(M)} \partial_M \lambda^M$, which is necessary for invariance of the hypermultiplet  action  \cite{Buchbinder:2022kzl}.
Thus, by introducing the analytic prepotentials $h^{++M}$, it became possible to covariantize the free hypermultiplet action  with respect to the super-diffeomorphisms \eqref{eqL super diff}.

To find out the physical field contents of the prepotentials introduced, we impose the Wess-Zumino type gauge:
\begin{equation}\label{eq: WZ gauge}
    \begin{split}
        &h_{WZ}^{++\alpha\dot{\alpha}} = -4i \theta^{+\beta}\bar{\theta}^{+\dot{\beta}} \Phi_{\beta\dot{\beta}}^{\alpha\dot{\alpha}}
        +
        (\bar{\theta}^+)^2 \theta^{+\beta} \psi_\beta^{\alpha\dot{\alpha}i} u^-_i
        -
        (\theta^+)^2 \bar{\theta}^{+\dot{\beta}}\bar{\psi}_{\dot{\beta}}^{\alpha\dot{\alpha}i}u^-_i
        +
        (\theta^+)^4 V^{\alpha\dot{\alpha}(ij)} u^-_i u^-_j
        \\
        &h_{WZ}^{++5} = -4i \theta^{+\beta}\bar{\theta}^{+\dot{\beta}} C_{\beta\dot{\beta}}
        +
        (\bar{\theta}^+)^2 \theta^{+\beta} \rho_\beta^{i} u^-_i
        -
        (\theta^+)^2 \bar{\theta}^{+\dot{\beta}}\bar{\rho}_{\dot{\beta}}^{i}u^-_i
        +
        (\theta^+)^4 S^{(ij)} u^-_i u^-_j
        \\
        &
        h_{WZ}^{++\alpha+} =  (\bar{\theta}^+)^2 \theta^+_{\beta} T^{(\alpha\beta)}
        +
        (\bar{\theta}^+)^2 \theta^{+\alpha}T
        +
        (\theta^+)^2 \bar{\theta}^+_{\dot{\beta}} P^{\alpha\dot{\beta}}
        +
        (\theta^+)^4 \chi^{\alpha i }u^-_i,
        \\
        & h_{WZ}^{++\dot{\alpha}+} = %\widetilde{\left(h_{WZ}^{++\alpha+}  \right)}
        %    =
        (\theta^+)^2 \bar{\theta}^+_{\dot{\beta}} \bar{T}^{(\dot{\alpha}\dot{\beta})} + (\theta^+)^2 \bar{\theta}^{+\dot{\alpha}} \bar{T}
        -
        (\bar{\theta}^+)^2 \theta^+_\beta \bar{P}^{\beta \dot{\alpha}}
        +
        (\theta^+)^4 \bar{\chi}^{\dot{\alpha}i}u^-_i.
    \end{split}
\end{equation}

In the linearized limit (the leading order in $\kappa_2$)  the fields in WZ gauge transform under the residual gauge freedom as \cite{Ivanov:2024gjo}:
    \begin{equation}\label{eq: gauge freedom Einstein SG}
        \begin{split}
        &\delta_\lambda \Phi^{\alpha\dot{\alpha}}_{\beta\dot{\beta}} \sim  \partial_{\beta\dot{\beta}} a^{\alpha\dot{\alpha}}
        -
         l^{\;\;\alpha)}_{(\beta} \delta^{\dot{\alpha}}_{\dot{\beta}}
        -
        \delta^{\alpha}_\beta l^{\;\;\dot{\alpha})}_{(\dot{\beta}},
    \\
    &
    \delta_\lambda P^{\alpha\dot{\alpha}} \sim  i \partial^{\dot{\alpha}\beta} l_{(\beta}^{\;\;\alpha)},
        \quad
                    \delta_\lambda C_{\alpha\dot{\alpha}} \sim  \partial_{\alpha\dot{\alpha}} c,
                    \\
        &\delta_\lambda \psi_\beta^{\alpha\dot{\alpha}i}
        \sim
        \partial^{\dot{\alpha}}_\beta \epsilon^{\alpha i},
    \quad
        \delta_\lambda \chi^{\alpha i} \sim   \Box \epsilon^{\alpha i},
    \quad
        \delta_\lambda \rho_\alpha^i \sim \partial_{\alpha\dot{\alpha}} \bar{\epsilon}^{\dot{\alpha}i}.
        \end{split}
    \end{equation}
So one can identify$\Phi^{\alpha\dot{\alpha}}_{\beta\dot{\beta}}$ with the gravitational vielbein, $\psi_\beta^{\alpha\dot{\alpha} i}$ with the doublet of gravitino fields, $C_{\alpha\dot{\alpha}}$ with the graviphoton.
Using local Lorentz transformation,one can impose the symmetric gauge
$\Phi_{\alpha\beta\dot{\alpha}\dot{\beta}} = \Phi_{(\alpha\beta)(\dot{\alpha}\dot{\beta})} + \epsilon_{\alpha\beta} \epsilon_{\dot{\alpha}\dot{\beta}} \Phi$.
All other gauge fields $(P^{\alpha\dot{\alpha}}, \chi^{ i }_\alpha, \rho^i_\alpha)$ can be redefined in terms of  physical fields so as to become invariant under the residual gauge freedom,
see \cite{Ivanov:2024gjo} for details. The ultimate $\mathcal{N}=2$ supermultiplet contains $\mathbf{40_B}+\mathbf{40_F}$ off-shell degrees of freedom:
\begin{equation}\label{eq: Einstein multiplet}
    \begin{split}
        \text{Physical fields:}&
        \qquad
        \left\{\Phi_{(\alpha\beta)(\dot{\alpha}\dot{\beta})}, \Phi \right\},
        \;\psi^i_{\alpha\beta\dot{\alpha}},
        \;
        C_{\alpha\dot{\alpha}},
        \\
        \text{Auxiliary fields:}&
        \qquad
        T, T^{(\alpha\beta)}, P^{\alpha\dot{\alpha}}, S^{(ij)}, V^{(ij)}_{\alpha\dot{\alpha}},
        \rho^i_\alpha, \chi^i_\alpha.
    \end{split}
\end{equation}
The off-shell component
content of this version of $\mathcal{N}=2$ Einstein supergravity was firstly found by Fradkin and
Vasiliev \cite{Fradkin:1979cw, Fradkin:1979as} and, independently, by de Wit and van Holten \cite{deWit:1979xpv}.
Thus, the harmonic superspace approach allows one to  obtain a $\mathcal{N}=2$ Einstein supergravity multiplet from the analytic gauging of supersymmetry.

In order to construct the superfield action it is useful to introduce superfields $H^{++M}$ and supersymmetry-invariant superfields $G^{++M}$ as vielbeins
in the new expansion of $\mathfrak{D}^{++}$:
\begin{equation}
    \mathfrak{D}^{++} = \partial^{++} + \kappa_2 H^{++M} \partial_M = \mathcal{D}^{++} + \kappa_2 G^{++M} \mathcal{D}_M.
\end{equation}
The supersymmetry-invariant derivatives $\mathcal{D}_M$ are  defined in eqs. \eqref{eq: spinor cov} of Appendix. The invariant $\mathcal{N}=2$ Einstein supergravity action was constructed in \cite{Galperin:1987em}:
\begin{equation}\label{eq: non-linear compensator}
    S^{full}_{sugra} = - \int d^4x d^8\theta du \, E H^{++5} H^{--5}.
\end{equation}
Here $E$ is $\mathcal{N}=2$ superspace measure constructed out of $h^{++M}$, see \cite{18, Galperin:1987em, Ivanov:2022vwc}. It has a complicated structure and will not be given here.
The vielbein coefficient $H^{--5}$ is a solution of the zero curvature equation $\mathfrak{D}^{++}H^{--5} = \mathfrak{D}^{--}H^{++5}$, with the derivative $\mathcal{D}^{--}$ being
defined from the curved zero-curvature condition  $[\mathfrak{D}^{++}, \mathfrak{D}^{--}] = \mathcal{D}^0$.

Since in the next sections we will be interested in the free (linearized) actions for $\mathcal{N}=2$ higher spins, it is instructive to give here the linearized action
of $\mathcal{N}=2$ Einstein supergravity \cite{Buchbinder:2021ite, Zupnik:1998td}:
\begin{equation}\label{eq: lin sugra}
    S^{lin}_{sugra} = - \int d^4x d^8 \theta du \; \left\{ G^{++\alpha\dot{\alpha}} G^{--}_{\alpha\dot{\alpha}} + 4 G^{++5} G^{--5}  \right\},
\end{equation}
where we used the supersymmetry-invariant superfields $G^{\pm\pm M}$. The negatively charged potentials can be found from
the flat zero-curvature equations  $\mathcal{D}^{++}G^{--M} = \mathcal{D}^{--}G^{++M}$. At the component level, the action \eqref{eq: lin sugra} is reduced to
the sum of the Pauli-Fiertz, Rarita-Schwinger and Maxwell actions, for details see \cite{Ivanov:2024gjo}.

\medskip
\subsubruninhead{$\mathcal{N}=2$ conformal supergravity}
\medskip

Although $\mathcal{N}=2$ Einstein supergravity can be formulated as described above, a more convenient and universal way of its construction consists in using the multiplet
of conformal supergravity in junction with the appropriate compensator multiplets, see e.g. \cite{Freedman:2012zz}.
The idea of this approach is to start from the interactions of superconformal gravity with superconformal matter multiplets and then to fix the conformal gauge freedom
by choosing non-trivial vacuum values for some matter fields.
Here is the simplest illustration of the conformal compensator method:
    \begin{equation}\label{eq: N=0 example}
    S[\phi, g_{mn}] = \int d^4x \sqrt{-g} \; \phi \left( \Box + \frac{1}{6} R \right) \phi
    \quad
    \xrightarrow{\phi = const}
    \quad
    S[g_{mn}]
    =
    \frac{1}{6} \int d^4 x \sqrt{-g}\, R.
\end{equation}
The most important motivation for us is that such a method of reproducing the gravity theory allows a  generalization to  $\mathcal{N}=2$ higher-spin  supergravity.

 \smallskip

The prepotentials of $\mathcal{N}=2$ conformal supergravity can be obtained in a way similar to that for $\mathcal{N}=2$ Einstein supergravity. In the conformal case one should gauge
the conformal transformations \eqref{eq: Lambda sc}:
\begin{equation}
    \hat{\Lambda}_{sc} \quad \to \quad \hat{\Lambda}_{sc}^{loc} = \lambda^{\alpha\dot{\alpha}}(\zeta) \partial_{\alpha\dot{\alpha}}
    +
    \lambda^{\hat{\alpha}+} (\zeta) \partial^-_{\hat{\alpha}}
    +
    \lambda^{\hat{\alpha}-}(\zeta, \theta^-) \partial^+_{\hat{\alpha}}
    +
    \lambda^{++} (\zeta) \partial^{--}.
\end{equation}
Since in this case  $[\mathcal{D}^{++}, \hat{\Lambda}^{loc}_{sc}] \neq \lambda^{++} \mathcal{D}^0$ (compare with \eqref{eq: N=2 sc}), the appropriate
covariantization of the harmonic derivative takes the form:
\begin{equation}\label{eq: cov der SC}
    \mathcal{D}^{++} \quad \to \quad \mathfrak{D}^{++} = \mathcal{D}^{++} + \kappa_2 \hat{\mathbb{H}}^{++}_{(s=2)},
\end{equation}
where the differential operator
\begin{equation}\label{eq: diff op conf SG}
    \hat{\mathbb{H}}^{++}_{(s=2)}
    =
    h^{++\alpha\dot{\alpha}}(\zeta) \partial_{\alpha\dot{\alpha}}
    +
    h^{++\hat{\alpha}+} (\zeta)\partial^{-}_{\hat{\alpha}}
    +
    h^{++\hat{\alpha}-} (\zeta, \theta^-) \partial^+_{\hat{\alpha}}
    +
    h^{(+4)} (\zeta) \partial^{--}.
\end{equation}
exhibits the transformation law
\begin{equation}\label{eq: delta H sc}
    \delta_\lambda  \hat{\mathbb{H}}_{(s=2)}^{++} = [\mathcal{D}^{++} + \kappa_2\hat{\mathbb{H}}_{(s=2)}^{++} , \hat{\Lambda}_{sc}^{loc}] - \lambda^{++} \mathcal{D}^0.
\end{equation}
It provides $\delta_\lambda \mathfrak{D}^{++} = \kappa_2 [\hat{\Lambda}^{sc}_{susy}, \mathfrak{D}^{++}] + \kappa_2 \delta_\lambda \hat{\mathbb{H}}_{(s=2)}^{++}   = - \kappa_2 \lambda^{++} \mathcal{D}^0$.
Note that in \eqref{eq: diff op conf SG} there is no derivative with respect to the auxiliary central charge coordinate $x^5$ (which breaks conformal symmetry),
but an additional harmonic derivative $\partial^{--}$ with the new vielbein coefficient $h^{(+4)}$ appear.

The non-analytic prepotential $h^{++\hat{\alpha}-}$ transforms according to
\begin{equation}
    \delta h^{++\hat{\alpha}-} = \left(\mathcal{D}^{++} +\kappa_2 \hat{\mathbb{H}}_{(s=2)}^{++}\right) \lambda^{-\hat{\alpha}}
    -
    \kappa_2 \hat{\Lambda}_{sc}^{loc} h^{++\hat{\alpha}-}
    -
    \lambda^{+\hat{\alpha}} + \lambda^{++}\theta^{-\hat{\alpha}},
\end{equation}
where $\lambda^{-\hat{\alpha}}$ is non-analytic. So like in $\mathcal{N}=2$ Einstein supergravity,  one can impose the analytic gauge $ h^{++\hat{\alpha}-} = 0$.

Using the harmonic covariant derivative \eqref{eq: cov der SC}, we specify the  interaction with the massless hypermultiplet:
\begin{equation}\label{eq: hyp sc}
    S_{hyp}^{sc} = - \frac{1}{2}  \int d\zeta^{(-4)} q^{+a} \left( \mathcal{D}^{++} + \kappa_2 \hat{\mathbb{H}}^{++}_{(s=2)} \right) q^+_a.
\end{equation}
This action is invariant under the gauge transformation \eqref{eq: delta H sc} accompanied by the hypermultiplet transformation:
\begin{equation}
    \delta_\lambda q^+_a = - \kappa_2 \hat{\mathcal{U}}_{sc} q^+_a  = - \kappa_2 \hat{\Lambda}_{sc}^{loc}q^+_a  - \frac{\kappa_2}{2} \Omega^{loc}_{sc}  q^+_a, \qquad
    \Omega_{sc}^{loc} = (-1)^{P(M)} \partial_M \lambda^M.
\end{equation}

In order to be convinced that the introduced gauge superfields describe the multiplet of $\mathcal{N}=2$ conformal supergravity,
one should impose Wess-Zumino type gauge:
\begin{equation}
    \begin{cases}
        h^{++\alpha\dot{\alpha}}
        =
        -4i \theta^{+\beta}\bar{\theta}^{+\dot{\beta}} \Phi^{\alpha\dot{\alpha}}_{\beta\dot{\beta}}
        +  (\bar{\theta}^+)^2 \theta^{+\beta} \psi^{\alpha\dot{\alpha}i}_\beta u^-_i + (\theta^+)^2 \bar{\theta}^{+\dot{\rho}} \bar{\psi}^{\alpha\dot{\alpha}i}_{\dot{\rho}}u_i^-
        \\\qquad\qquad\qquad\qquad\qquad\qquad\qquad\qquad\qquad  +  (\theta^+)^4 V^{\alpha\dot{\alpha}(ij)}u^-_iu^-_j\,,\\
        h^{++\alpha+} =(\theta^+)^2 \bar{\theta}^+_{\dot{\alpha}} P^{\alpha\dot{\alpha}}
        + \left(\bar{\theta}^+\right)^2 \theta^+_\beta T^{(\alpha\beta)}
        +  (\theta^+)^4 \chi^{\alpha i}u^-_i, \;\;\;
        \\
        h^{++\dot{\alpha}+} = \widetilde{h^{++\alpha+}}\,,
        \\
        h^{(+4)} \;\;\,= (\theta^+)^4 D\,.
    \end{cases}
\end{equation}
The remaining fields are determined up to the residual gauge transformations which differ from \eqref{eq: gauge freedom Einstein SG}:
\begin{equation}\label{eq: conf grav gauge}
    \begin{split}
    &\delta_\lambda  \Phi^{\alpha\dot{\alpha}}_{\beta\dot{\beta}}
    \sim
    \partial_{\beta\dot{\beta}} a^{\alpha\dot{\alpha}}
    -
    l^{\;\;\alpha)}_{(\beta} \delta^{\dot{\alpha}}_{\dot{\beta}}
    -
    \delta^{\alpha}_\beta l^{\;\;\dot{\alpha})}_{(\dot{\beta}}
    + d     \delta^{\alpha}_\beta\delta^{\dot{\alpha}}_{\dot{\beta}},
    \\
    &
    \delta_\lambda \psi^{\alpha\dot{\alpha}i}_\beta
        \sim
    \partial_\beta^{\dot{\alpha}} \epsilon^{\alpha i} + \delta^\alpha_\beta  \bar{\eta}^{\dot{\alpha} i},
    \quad
    \delta_\lambda \chi^{\alpha i}  \sim \partial^{\alpha\dot{\alpha}} \bar{\eta}_{\dot{\alpha}}^i
    +
    \Box \epsilon^{\alpha i},
    \\
    &
    \delta_\lambda V_{\alpha\dot{\alpha}}^{(ij)}    \sim \partial_{\alpha\dot{\alpha}} v^{(ij)},
    \quad
    \delta_\lambda P_{\alpha\dot{\alpha}}   \sim \partial_{\alpha\dot{\alpha}} (r + id) -i \partial^{\dot{\alpha}\beta} l_{(\beta}^{\;\alpha)} - i k_{\alpha\dot{\alpha}},
    \\
    &
    \delta_\lambda D    \sim \partial^{\alpha\dot{\alpha}} k_{\alpha\dot{\alpha}}.
    \end{split}
\end{equation}

As compared to the  transformations \eqref{eq: gauge freedom Einstein SG}, here one finds additional gauge parameters corresponding to the local dilatations ($d$), local special conformal transformations
($k_{\alpha\dot{\alpha}}$), local conformal supersymmetry ($\eta^{i}_{\alpha}$, $\bar{\eta}^i_{\dot{\alpha}}$), local $U(1)_R$ ($r$) and local $SU(2)_R$ ($v^{(ij)}$) transformations.
Using this freedom, one can gauge away the trace of graviton,  the trace of the spin $\frac{3}{2}$ field and the imaginary part of $P_{\alpha\dot{\alpha}}$. The resulting field representation is constituted by
\begin{equation}\label{eq: Conf mult}
    \begin{split}
    \text{Gauge fields:}&
    \qquad
    \Phi_{(\alpha\beta)(\dot{\alpha}\dot{\beta})},
    \;\psi^i_{(\alpha\beta)\dot{\alpha}},
    \;
    V^{(ij)}_{\alpha\dot{\alpha}}, \;   R_{\alpha\dot{\alpha}} := Re(P_{\alpha\dot{\alpha}}),
    \\
    \text{Non-gauge fields:}&
    \qquad
  \chi_\alpha^i, \; T^{(\alpha\beta)}, \; D,
    \end{split}
\end{equation}
that is total of $\mathbf{24}_B + \mathbf{24}_F$ off-shell degrees of freedom inherent to $\mathcal{N}=2$ Weyl multiplet   \cite{ deWit:1979dzm, Bergshoeff:1980is, deWit:1980lyi}.
Thus, we have shown that the analytic superfields $h^{++M}$ can indeed be identified with the unconstrained prepotentials of $\mathcal{N}=2$ conformal supergravity.

\medskip

The linearized action of conformal ${\cal N}=2$ supergravity can be constructed by generalizing the approach to the linearized higher-derivative invariants described in \cite{Ivanov:2024gjo},
\begin{equation}\label{eq: Weyl action}
    S_{Weyl} = \int d^4x d^4\theta \,   \mathcal{W}^{(\alpha\beta)}     \mathcal{W}_{(\alpha\beta)} + c.c.
\end{equation}
Here we have introduced the linearized $\mathcal{N}=2$ super-Weyl tensor\footnote{The supersymmetry-invariant potentials are introduced
as $\hat{\mathbb{H}}_{(s=2)} = G^{++M} \mathcal{D}_M$, $\mathcal{D}^{--}G^{++M} = \mathcal{D}^{++}G^{--M} $, $\mathcal{D}^{++} G^{---}_{\hat{\beta}} = G^{--+}_{\hat{\beta}}$.}
\begin{equation}
    \mathcal{W}_{(\alpha\beta)} =\, (\bar{\mathcal{D}}^+)^2
    \left(  \partial_{(\alpha}^{\dot{\beta}} G^{--}_{\beta)\dot{\beta}}
    -
    \mathcal{D}^-_{(\alpha} G^{--+}_{\beta)} + \mathcal{D}^+_{(\alpha} G^{---}_{\beta)} \right),
\end{equation}
which satisfy the chirality and the harmonic independence conditions,
\begin{equation} \bar{\mathcal{D}}^{\pm}_{\dot{\rho}}   \mathcal{W}_{(\alpha\beta)}  = 0,
\qquad \mathcal{D}^{\pm\pm} \mathcal{W}_{(\alpha\beta)}  = 0\,,
\end{equation}
and is invariant under gauge transformations to the leading order.

In  components, the $\mathcal{N}=2$ super Weyl tensor is represented as
\begin{equation}
    \begin{split}
        \mathcal{W}_{(\alpha\beta)}
        \sim &\; \theta^{+\rho}     \hat{C}^i_{(\alpha\beta\rho)}u^-_i  - \theta^{-\rho}    \hat{C}^i_{(\alpha\beta\rho)}u^+_i +   \theta^{+(\rho} \theta^{-\sigma)} C_{(\alpha\beta\rho\sigma)}
        +
        \theta^+_{(\alpha} \theta^{-\rho)} R_{\rho \beta}
        \\
        & \; +(\theta^+)^2 V_{(\alpha\beta)}^{(ij)} u^-_i u^-_j
        -
        2(\theta^+ \theta^-) V_{(\alpha\beta)}^{(ij)} u^+_i u^-_j
        +
        (\theta^-)^2 V_{(\alpha\beta)}^{(ij)} u^+_i u^+_j
        \\
        & \; + (\theta^+)^2 \theta^{-\rho} \check{\bar{C}}^i_{(\alpha\beta\rho)} u^-_i
        -
        (\theta^-)^2  \theta^{+\rho} \check{\bar{C}}^i_{(\alpha\beta\rho)} u^+_i + \dots,
    \end{split}
\end{equation}
where we have introduced the linearized gauge-invariant tensors for conformal graviton, for doublet of conformal gravitino and for $R$-symmetry gauge fields:
\begin{subequations}
\begin{equation}
    C_{(\alpha\beta\rho\sigma)} = \partial_{(\alpha}^{\dot{\rho}} \partial_\beta^{\dot{\sigma}} \Phi_{\rho\sigma) (\dot{\rho}\dot{\sigma})},
    \quad
    V_{(\alpha\beta)}^{(ij)} = \partial_{(\alpha}^{\dot{\alpha}} V_{\alpha\dot{\alpha}}^{(ij)},
    \quad
    R_{(\alpha\beta)} = \partial_{(\alpha}^{\dot{\rho}} R_{\beta)\dot{\rho}}\,,
\end{equation}
\begin{equation}
    \hat{C}^i_{(\alpha\beta\rho)} = \partial^{\dot{\rho}}_{(\rho} \psi^i_{\alpha\beta)\dot{\rho}},
    \qquad
    \check{\bar{C}}^i_{(\alpha\beta\rho)} = \partial^{\dot{\rho}}_{(\alpha} \partial^{\dot{\sigma}}_{\beta} \bar{\psi}^i_{\rho) (\dot{\rho}\dot{\sigma})}.
\end{equation}
\end{subequations}
So the component linearized form of Weyl action \eqref{eq: Weyl action} is written as
\begin{equation}
    S_{Weyl} \sim \int d^4x \left\{ C^{\alpha(4)}C_{\alpha(4)} + \hat{C}^{\alpha(3)}_i  \check{\bar{C}}^i_{\alpha(3)} + V^{\alpha(2)}_{(ij)} V_{\alpha(2)}^{(ij)} + R^{(\alpha\beta)}R_{(\alpha\beta)} \right\}.
\end{equation}
This action contains higher derivatives.

The complete nonlinear action of $\mathcal{N}=2$ conformal supergravity in harmonic superspace is as yet unknown,
although it has been constructed within other approaches, see  \cite{Bergshoeff:1980is, Kuzenko:2015jxa, Kuzenko:2022ajd}.

\medskip
\subsubruninhead{$\mathcal{N}=2$ superconformal compensators}
\medskip

The comparison of  the multiplets \eqref{eq: Einstein multiplet}, \eqref{eq: Conf mult} with the relevant gauge transformations \eqref{eq: gauge freedom Einstein SG}, \eqref{eq: conf grav gauge}
 shows that, nullifying the gauge ``conformal'' parameters $d, \eta^i_\alpha, k_{\alpha\dot{\alpha}}, v^{(ij)}$ in $\mathcal{N}=2$ conformal supergravity gauge group,
 we can restore the physical fields of $\mathcal{N}=2$ Einstein supergravity. The natural way to do this is to consider the interaction of $\mathcal{N}=2$ conformal supergravity with matter conformal supermultiplets,
 for example with the hypermultiplet as in \eqref{eq: hyp sc}.  Then the ``conformal'' gauge parameters can be fixed by imposing the suitable set of gauge conditions corresponding to the choice
 of vacuums that violate these symmetries (recall the example \eqref{eq: N=0 example}).

For $\mathcal{N}=2$ supergravity one needs two compensators \cite{deWit:1980lyi}: $\mathcal{N}=2$  vector multiplet and a matter multiplet. For the latter there exist four possible
options:
% as a superconformal multiplet of $\mathcal{N}=2$ matter:
\begin{itemize}
    \item Hypermultiplet

    \item Non-linear multiplet

    \item Improved tensor multiplet

    \item Hypermultiplet with non-trivial central charge.
\end{itemize}

A detailed discussion of these versions of $\mathcal{N}=2$ supergravity and additional references can be found in \cite{18, Galperin:1987ek, Ivanov:2022vwc, Galperin:1985tn}.
In particular, $\mathcal{N}=2$ Einstein supergravity  action \eqref{eq: non-linear compensator} corresponds to the choice of non-linear $\mathcal{N}=2$ multiplet as a compensator.

Schematically, the structure of the Lagrangians of such theories can be represented as:
\begin{equation}
    S_{sg} \sim - S^{conf}_{vec} - S^{conf}_{matter},
\end{equation}
where the corresponding actions are covariantized with respect to the  $\mathcal{N}=2$ conformal supergravity gauge group and are taken with the wrong signs (which is typical for compensators).

Note that in the linearized approximation we considered the action \eqref{eq: lin sugra} corresponding to the choice
of  non-linear compensator. It will be interesting to construct harmonic superspace actions for other versions of $\mathcal{N}=2$ linearized supergravity.

\section{$\mathcal{N}=2$ higher spin supermultiplets}
\label{eq: sec HS}

The basic difference between the harmonic approach \cite{Ivanov:2022vwc} and other superfield approaches to $\mathcal{N}=2$ supergravity \cite{Kuzenko:2022ajd}
is that the harmonic formulation naturally yields unconstrained prepotentials, the feature which is absent in other approaches.
It is highly non-trivial that the harmonic formulation allows a straightforward generalization of $\mathcal{N}=2$ supergravity prepotentials to $\mathcal{N}=2$  higher spins.

In this section, we show how $\mathcal{N}=2$ supermultiplets of higher spins appear in the harmonic approach, construct their cubic vertices of interaction with the hypermultiplet
(which is consistent to the leading order) and present the linearized $\mathcal{N}=2$ gauge-invariant actions.

\subsubruninhead{$\mathcal{N}=2$ Einstein-Fronsdal  higher spins}

\medskip

Similar to the case of $\mathcal{N}=2$ supergravity, we gauge the higher-spin supersymmetry transformations  \eqref{eq: spin s susy}:
\begin{equation}
    \hat{\Lambda}_{susy}^{(s)}
    \quad
    \to
    \quad
    \hat{\Lambda}^{(s)}
    =
    \hat{\Lambda}^{\alpha(s-2)\dot{\alpha}}  \partial^{s-2}_{\alpha(s-2)\dot{\alpha}(s-2)},
\end{equation}
where
    \begin{equation}\label{Lambda rigid}
    \hat{\Lambda}^{\alpha(s-2)\dot{\alpha}(s-2)} := \lambda^{\alpha(s-2)\dot{\alpha}(s-2)M} \partial_M,
    \qquad M = (\alpha\dot{\alpha}, \alpha+, \dot{\alpha}+, 5).
\end{equation}
Since $[\mathcal{D}^{++}, \hat{\Lambda}^{(s)}] \neq 0$, we define the covariant harmonic derivative
\begin{equation}\label{eq: spin s cov}
    \mathcal{D}^{++}
    \quad
    \to
    \quad
    \mathfrak{D}^{++}_{(s)}
    =
    \mathcal{D}^{++} + \kappa_s \hat{\mathcal{H}}^{++}_{(s)},
\end{equation}
where the spin $\mathbf{s}$ analytic differential operator is defined as
\begin{equation}\label{G-operator}
    \hat{\mathcal{H}}^{++}_{(s)} := h^{++\alpha(s-2)\dot{\alpha}(s-2)M} \partial_{M}\partial^{s-2}_{\alpha(s-2)\dot{\alpha}(s-2)},
    \qquad
    \partial_M = \{\partial_{\alpha\dot{\alpha}}, \partial^-_\alpha, \partial^-_{\dot{\alpha}}, \partial_5\}.
\end{equation}

The invariance of the harmonic derivative $\mathfrak{D}^{++}_{(s)}$, in contrast to $\mathcal{N}=2$ supergravity, can be achieved only in the leading order in $\kappa_s$:
\begin{equation}
    \delta_\lambda \mathfrak{D}^{++}_{(s)}
    =
    \kappa_s [\hat{\Lambda}^{(s)}, \mathcal{D}^{++}+ \kappa_s \hat{\mathcal{H}}^{++}_{(s)}]
    +
    \kappa_s \delta_\lambda  \hat{\mathcal{H}}^{++}_{(s)}
    =
    0
    +
    \mathcal{O}(\kappa_s^2).
\end{equation}
This defines the linearized transformation law  of the operator \eqref{G-operator}:
\begin{equation}\label{eq: spin s gauge}
    \delta_\lambda \hat{\mathcal{H}}^{++}_{(s)} = [\mathcal{D}^{++}, \hat{\Lambda}^{\alpha(s-2)\dot{\alpha}}] \partial^{s-2}_{\alpha(s-2)\dot{\alpha}(s-2)}.
\end{equation}
Invariance in the following orders requires the introduction of a new type of prepotentials, which are not reduced to $\hat{\mathcal{H}}^{++}_{(s)}$.
This is also because of the property that the relevant transformations (\eqref{eq: local spin s} below) do not form a closed algebra.
Thus we meet obstacles to constructing the  consistent nonlinear$\mathcal{N}=2$ higher-spin actions.

Using the covariant derivative \eqref{eq: spin s cov}, one can construct the coupling consistent to the leading order:
\begin{equation}\label{action-final}
    S^{(s)}_{hyp}
    =
    - \frac{1}{2} \int d\zeta^{(-4)}\;
    q^{+a} \left(\mathcal{D}^{++} +  \kappa_s \hat{\mathcal{H}}^{++}_{(s)}  (J)^{P(s)}  \right) q^+_a\,.
\end{equation}
 The corresponding hypermultiplet transformation has the form \cite{Buchbinder:2022kzl}:
    \begin{equation}\label{transformations}
    \delta^{(s)}_{\lambda} q^{+a}
    = - \kappa_s    \hat{\mathcal{U}}^{(s)}_{susy} q^{+a},
\end{equation}
where
\begin{equation}\label{eq: local spin s}
    \begin{split}
    \hat{\mathcal{U}}^{(s)}_{susy}
    =&
    - \frac{1}{2}\left\{\hat{\Lambda}^{\alpha(s-2)\dot{\alpha}(s-2)}, \partial^{s-2}_{\alpha(s-2)\dot{\alpha}(s-2)}\right\} (J)^{P(s)}
    \\&
    -\frac{1}{2}
    \partial^{s-2}_{\alpha(s-2)\dot{\alpha}(s-2)} \Omega^{\alpha(s-2)\dot{\alpha}(s-2)} (J)^{P(s)}\,,
    \end{split}
\end{equation}
\begin{equation}
    \Omega^{\alpha(s-2)\dot{\alpha}(s-2)} := (-1)^{P(M)} \partial_M \lambda^{M\alpha(s-2)\dot{\alpha}(s-2)}.
\end{equation}

In order to exhibit the physical contents of the  superfields introduced, we can make use of the gauge freedom \eqref{eq: spin s gauge} to impose the Wess-Zumino type gauge:
\begin{equation}\label{eq: WZ gauge}
    \begin{split}
        &  h_{WZ}^{++\alpha(s-1)\dot{\alpha}(s-1)}
        =
        -4i \theta^{+\beta} \bar{\theta}^{+\dot{\beta}} \Phi_{\beta\dot{\beta}}^{\alpha(s-1)\dot{\alpha}(s-1)}
        \\
        &\qquad\qquad\qquad\qquad
        +  (\bar{\theta}^+)^2 \theta^{+\beta} \psi_{\beta}^{\alpha(s-1)\dot{\alpha}(s-1)i}u^-_i
        +\, (\theta^+)^2 \bar{\theta}^{+\dot{\beta}} \bar{\psi}_{\dot{\beta}}^{\alpha(s-1)\dot{\alpha}(s-1)i}u_i^-
        \\&\qquad\qquad\qquad\qquad
        +  (\theta^+)^2 (\bar{\theta}^+)^2 V^{\alpha(s-1)\dot{\alpha}(s-1)(ij)}u^-_iu^-_j\,,
        \\
        &  h_{WZ}^{++\alpha(s-2)\dot{\alpha}(s-2)} =
        -4i \theta^{+\beta} \bar{\theta}^{+\dot{\beta}} C^{\alpha(s-2)\dot{\alpha}(s-2)}_{\beta\dot{\beta}}
        \\&\qquad\qquad\qquad\qquad+ (\bar{\theta}^+)^2 \theta^{+\beta} \rho_\beta^{\alpha(s-2)\dot{\alpha}(s-2)i}u^-_i
        + (\theta^+)^2 \bar{\theta}^{+\dot{\beta}} \bar{\rho}_{\dot{\beta}}^{\alpha(s-2)\dot{\alpha}(s-2)i}u_i^-
        \\&\qquad\qquad\qquad\qquad+ (\theta^+)^2 (\bar{\theta}^+)^2 S^{\alpha(s-2)\dot{\alpha}(s-2)(ij)}u^-_iu^-_j\,,
        \\
        &  h_{WZ}^{++\alpha(s-1)\dot{\alpha}(s-2)+} = (\theta^+)^2 \bar{\theta}^+_{\dot{\beta}} P^{\alpha(s-1)\dot{\alpha}(s-2)\dot{\beta}}
        +  \left(\bar{\theta}^+\right)^2 \theta^+_\beta T^{\dot\alpha(s-2)\alpha(s-1)\beta}
        \\&\qquad\qquad\qquad\qquad+  (\theta^+)^2 (\bar{\theta}^+)^2 \chi^{\alpha(s-1)\dot{\alpha}(s-2)i}u^-_i\,,
        \\
        &  h_{WZ}^{++\dot{\alpha}(s-1)\alpha(s-2)+} = \widetilde{\left(h_{WZ}^{++\alpha(s-1)\dot{\alpha}(s-2)+}\right)}\,.
    \end{split}
\end{equation}
The WZ fields are defined up to the residual gauge freedom
\begin{subequations}\label{eq: res fronsdal HS}
\begin{equation}
    \delta_\lambda \Phi^{\alpha(s-1)\dot{\alpha}(s-1)}_{\beta\dot{\beta}} \sim \partial_{\beta\dot{\beta}} a^{\alpha(s-1)\dot{\alpha}(s-1)} - l_{(\beta}^{\alpha(s-1))(\dot{\alpha}(s-2)}\delta_{\dot{\beta}}^{\dot{\alpha})}
    -
    \bar{l}_{(\dot{\beta}}^{\dot{\alpha}(s-1))(\alpha(s-2)}\delta^{\alpha)}_\beta,
\end{equation}
\begin{equation}
    \delta_\lambda C^{\alpha(s-2)\dot{\alpha}(s-2)}_{\beta\dot{\beta}} \sim \partial_{\beta\dot{\beta}} a^{\alpha(s-2)\dot{\alpha}(s-2)} - n_{(\beta}^{\alpha(s-2))(\dot{\alpha}(s-3)}\delta_{\dot{\beta}}^{\dot{\alpha})}
    -
    \bar{n}_{(\dot{\beta}}^{\dot{\alpha}(s-2))(\alpha(s-3)}\delta^{\alpha)}_\beta,
\end{equation}
\begin{equation}
    \delta \psi_\beta^{\alpha(s-1)\dot{\alpha}(s-1)i} \sim \partial_\beta^{(\dot{\alpha}} \xi^{\alpha(s-1)\dot{\alpha}(s-1))i},
    \qquad
    \delta \bar{\rho}^{\alpha(s-2)(\dot{\alpha}(s-3)\dot{\beta})i}_{\dot{\beta}}
    \sim
    \partial_{\beta\dot{\beta}} \xi^{(\alpha(s-2)\beta)(\dot{\alpha}(s-3)\dot{\beta})i}.
\end{equation}
\end{subequations}
After fixing the local Lorentz invariance we come to the set of physical fields
\begin{equation}\label{eq: Fronsdal multiplet}
    \begin{split}
        &\left\{\Phi_{\alpha(s)\dot{\alpha}(s)},\Phi_{\alpha(s-2)\dot{\alpha}(s-2)} \right\},
        \;\left\{\psi_\beta^{\alpha(s-1)\dot{\alpha}(s-1)i}, \bar{\rho}^{\alpha(s-2)(\dot{\alpha}(s-3)\dot{\beta})i}_{\dot{\beta}} \right\},
        \\&
            \left\{C_{\alpha(s-1)\dot{\alpha}(s-1)},C_{\alpha(s-3)\dot{\alpha}(s-3)} \right\}.
            \end{split}
\end{equation}
They correspond to the Fronsdal spin $s$ gauge fields, the doublet of Fang-Fronsdal spin $s-\frac{1}{2}$ fermionic gauge fields and the Fronsdal spin $s-1$ gauge fields.
Other fields can be shown to be auxiliary after proper redefinitions which make them gauge-invariant (see details in \cite{Buchbinder:2021ite}). Finally, we
encounter total of  $\mathbf{8 [2s^2 -2s+1]_B} + \mathbf{8 [2s^2 -2s+1]_F}$ off-shell degrees of freedom.

The linearized gauge invariant actions have the universal form for all superspins:
\begin{equation}\label{eq: ActionsGen}
    \begin{split}
        & S_{(s)} = (-1)^{s+1} \int d^4x
        d^8\theta du \,\Big\{G^{++
            \alpha(s-1)\dot\alpha(s-1)}G^{--}_{\alpha(s-1)\dot\alpha(s-1)} \\
        &\;\;\;\;\;  \qquad\qquad \qquad\qquad \qquad\qquad+\,
        4G^{++\alpha(s-2)\dot\alpha(s-2)}G^{--}_{\alpha(s-2)\dot\alpha(s-2)}
        \Big\}\,,
    \end{split}
\end{equation}
where the supersymmetry-invariant superfields $G^{++M}$ are defined as vielbeins in
\begin{equation}
        \hat{\mathcal{H}}^{++}_{(s)} := G^{++\alpha(s-2)\dot{\alpha}(s-2)M} \mathcal{D}_{M}\partial^{s-2}_{\alpha(s-2)\dot{\alpha}(s-2)},
\end{equation}
and $G^{--M}$ as solutions of  the harmonic zero-curvature equations:
\begin{equation}
    \mathcal{D}^{++} G^{--M} = \mathcal{D}^{--} G^{++M}.
\end{equation}

In conclusion of this section we note that it is possible to construct a more general class of $\mathcal{N}=2$ higher-spin cubic vertices,
using the conserved $\mathcal{N}=2$ higher-spin supercurrents \cite{Zaigraev:2024ryg}. We will not touch this issue here.

\medskip
\subsubruninhead{$\mathcal{N}=2$ superconformal higher spins}
\medskip

In order to construct $\mathcal{N}=2$ superconformal multiplets of higher spins, we will follow a slightly different strategy,
which is more suitable for the superconformal case. We will start from  the higher-spin hypermultiplet vertex of general type and,
requiring invariance with respect to the superconformal transformations, determine the set of analytical potentials which are necessary
to secure this invariance.

The invariance under $\mathcal{N}=2$ superconformal transformations \eqref{eq: sc transf} requires including all possible types of the
superspace derivatives $s-1$, $s-3$, $\dots$:
\begin{equation}\label{eq: spin s sc vertex}
      S^{(s)}_{hyp-conf}
        =
        -
        \frac{1}{2} \int d\zeta^{(-4)}\,
        q^{+a} \left(\mathcal{D}^{++} + \kappa_s \hat{\mathbb{H}}^{++}_{(s)} (J)^{P(s)} \right) q^+_a\,,
\end{equation}
where we introduced the analytic differential operator of degree  $s-1$:
\begin{equation}\label{eq: sc spin s operator}
    \hat{\mathbb{H}}^{++}_{(s)} := h^{++M_1\dots M_{s-1}} \partial_{M_{s-1}} \dots \partial_{M_1}
    +
    h^{++M_1\dots M_{s-3}}  \partial_{M_{s-3}} \dots \partial_{M_1}
    +
    \dots\,,
\end{equation}
Then, defining  $\mathcal{N}=2$ superconformal transformations of the prepotentials as
\begin{equation}
    \delta_{sc} \hat{\mathbb{H}}^{++}_{(s)}  = [\hat{\mathbb{H}}^{++}_{(s)} , \hat{\Lambda}_{sc}] + \frac{1}{2} [\hat{\mathbb{H}}^{++}_{(s)} , \Omega],
\end{equation}
it is possible to ensure the $\mathcal{N}=2$ superconformal invariance of the vertex \eqref{eq: spin s sc vertex}.
It is worth noting that simultaneously one covariantizes the vertex \eqref{eq: spin s sc vertex} under the generic conformal supergravity transformations.

The vertex \eqref{eq: spin s sc vertex} is invariant under the  huge set of gauge transformations
\begin{equation}\label{eq: sc spin s transformations}
    \delta^{(s,k)}_\lambda  \hat{\mathbb{H}}^{++}_{(s)} =
    \frac{1}{2} \left[\mathcal{D}^{++}, \left\{ \hat{\Lambda}^{M_1\dots M_{k-2}}, \partial_{M_{k-2}}\dots \partial_{M_1} \right\}_{AGB} \right],
    \quad
    k=s,s-2, \dots\;,
\end{equation}
 \begin{equation}\label{eq: spin s gauge transformations - k-1}
    \begin{split}
        \delta_\lambda^{(k)} q^{+}_{a}
        =& - \kappa_s \,
        \hat{\mathcal{U}}^{(k)}_{s}   q^{+}_{a}
        \\
        =&
        - \frac{\kappa_s}{2} \left\{\hat{\Lambda}^{M_1 \dots M_{k-2}}, \partial_{M_{k-2}} \dots  \partial_{M_1} \right\}_{AGB} \left( J \right)^{P(s)} q^{+}_{a}
        \\
        &
        - \frac{\kappa_s}{4} \left\{\Omega^{M_1 \dots M_{k-2}}, \partial_{M_{k-2}} \dots  \partial_{M_1} \right\}_{AGB} \left( J \right)^{P(s)} q^{+}_{a}.
    \end{split}
\end{equation}
Different values of $k$ correspond to different spin contributions to the operator $\hat{\mathbb{H}}^{++}_{(s)} $.
The algebraic structure of these transformations will be sketched in the next section. Requiring $\delta^{(s,k)}_\lambda  \hat{\mathbb{H}}^{++}_{(s)} = 0$ gives rise to
the equations which yield rigid higher-spin $\mathcal{N}=2$ conformal  supersymmetries.

The superfield gauge transformations  \eqref{eq: sc spin s transformations} lead to the gauge transformations of the component fields
(for simplicity, we consider the  spin $s$ and spin $s-\frac{1}{2}$ fields and present the transformations in a schematic form):
\begin{subequations}\label{eq: conf spin s}
\begin{equation}
    \begin{split}
    \delta_\lambda \Phi^{\alpha(s-1)\dot{\alpha}(s-1)}_{\beta\dot{\beta}} \sim &\;
    \partial_{\beta\dot{\beta}} a^{\alpha(s-1)\dot{\alpha}(s-1)}
    \\
    &- l_{(\beta}^{\alpha(s-1))(\dot{\alpha}(s-2)}\delta_{\dot{\beta}}^{\dot{\alpha})}
    -
    \bar{l}_{(\dot{\beta}}^{\dot{\alpha}(s-1))(\alpha(s-2)}\delta^{\alpha)}_\beta
    \\&
    +
    \delta_\beta^{(\alpha} \delta_{\dot{\beta}}^{(\dot{\alpha}} d^{\alpha(s-2))\dot{\alpha}(s-2))},
    \end{split}
\end{equation}
\begin{equation}
    \delta_\lambda \psi_\beta^{\alpha(s-1)\dot{\alpha}(s-1)i}
    \sim
    \partial_\beta^{(\dot{\alpha}} \epsilon^{\alpha(s-1)\dot{\alpha}(s-2))i}
    +
    \delta_\beta^{(\alpha} \bar{\eta}^{\alpha(s-2))\dot{\alpha}(s-1)i}.
\end{equation}
\end{subequations}
It is instructive to compare these transformations with eqs. \eqref{eq: res fronsdal HS}. As in the supergravity case, when breaking
local higher-spin dilation and local conformal supersymmetry, we restore the gauge freedom of Fronsdal and Fang-Fronsdal fields.
This indicates a possible connection with theories of the Fronsdal type.

\smallskip

Using the higher-spin Lorentz, dilatation, and conformal supersymmetry parameters to gauge away traces  of fields in \eqref{eq: conf spin s},
we are finally left with the spin $s$ and $s-\frac{1}{2}$ Fradkin-Tseytlin conformal fields \cite{Fradkin:1985am}:
\begin{equation}
    \begin{split}
    \delta_\lambda \Phi_{\alpha(s)\dot{\alpha}(s)} &\sim \partial_{(\alpha(\dot{\alpha}} a_{\alpha(s-1))\dot{\alpha}(s-1))},
\\
    \delta_\lambda \psi^i_{\alpha(s)\dot{\alpha}(s-1)} & \sim \partial_{(\alpha(\dot{\alpha}} \epsilon^i_{\alpha(s-1))\dot{\alpha}(s-2))}.
    \end{split}
\end{equation}
The rest of the component fields can be tackled in a similar way (see the detailed analysis of the $s=3$ case in \cite{Buchbinder:2024pjm}).
Everything is reduced to the Fradkin-Tseytlin type higher-spin fields:
\begin{equation}\label{eq: conf WZ}
    %   \begin{cases}
        \begin{split}
            &h_{WZ}^{++\alpha(s-1)\dot{\alpha}(s-1)}
            =
            -4i \theta^{+}_{\rho}\bar{\theta}^{+}_{\dot{\rho}} \Phi^{(\rho\alpha(s-1))(\dot{\rho}\dot{\alpha}(s-1))}
            -
            (\bar{\theta}^+)^2 \theta^{+}_{\rho} \psi_{}^{(\rho\alpha(s-1))\dot{\alpha}(s-1)i} u_i^-
            \\&\;\;\;\;\qquad\qquad\qquad\quad-
            (\theta^+)^2 \bar{\theta}^{+}_{\dot{\rho}} \bar{\psi}^{\alpha(s-1)(\dot{\alpha}(s-1)\dot{\rho})i} u_i^-
            +
            (\theta^+)^4 V^{\alpha(s-1)\dot{\alpha}(s-1)ij}u^-_i u^-_j\,,\\
            &h_{WZ}^{++\alpha(s-1)\dot{\alpha}(s-2)+}
            =
            (\theta^{+})^2 \bar{\theta}^{+}_{\dot{\nu}} P^{\alpha(s-1)(\dot{\alpha}(s-2)\dot{\nu})}
            +
            (\bar{\theta}^{+})^2 \theta^{+}_{\nu} T^{(\alpha(s-1)\nu)\dot{\alpha}(s-2)}_{}
            \\&\qquad\qquad\qquad\qquad\qquad\quad\qquad\qquad\qquad\quad\;\;\,+
            (\theta^{+})^4 \chi^{\alpha(s-1)\dot{\alpha}(s-2)i}u_i^-\,,
            \\
            &h_{WZ}^{++\alpha(s-2)\dot{\alpha}(s-1)+}=
            \widetilde{h_{WZ}^{++\alpha(s-1)\dot{\alpha}(s-2)+}}\,,
            \\&h_{WZ}^{(+4)\alpha(s-2)\dot{\alpha}(s-2)} =
            (\theta^+)^4 D^{\alpha(s-2)\dot{\alpha}(s-2)}\,.
        \end{split}
        %   \end{cases}
\end{equation}
All other analytic prepotentials in \eqref{eq: sc spin s operator} play an auxiliary role: it is possible to impose a gauge
in which they are vanishing (just this gauge was chosen in eqs. \eqref{eq: conf WZ}).

It should be emphasized  that all fields in \eqref{eq: conf WZ} are gauge fields, and this is a crucial difference of the resulting supermultiplet from the multiplet
of $\mathcal{N}=2$ conformal supergravity ($\bf{s}=2$) and non-conformal $\mathcal{N}=2$ higher-spin multiplets.
This means that all fields appear in the action in a dynamical way, through terms containing derivatives.
The component decomposition of the spin $\mathbf{s}$ superconformal multiplet can be conveniently represented as:
\begin{equation*}
    s
    \quad \leftrightarrow \quad
    2 \times (s-1/2)
    \quad \leftrightarrow \quad
    4 \times (s-1)
    \quad \leftrightarrow \quad
    2 \times (s-3/2)
    \quad \leftrightarrow \quad
    (s-2)
\end{equation*}
plus the ``hook'' or conformal pseudo-spin $s-1$ gauge field  $T^{\alpha(s)\dot{\alpha}(s-2)}$.
We have total of $\mathbf{8(2s-1)_B+8(2s-1)_F}$ off-shell degrees of freedom.

The natural conjecture is that the free actions for $\mathcal{N}=2$ superconformal higher spins are constructed quite analogously to the action \eqref{eq: Weyl action} of $\mathcal{N}=2$ conformal supergravity:
\begin{equation}
    S^{(s)}_{Weyl} = \int d^4x d^4\theta \; \mathcal{W}^{\alpha(2s-2)} \mathcal{W}_{\alpha(2s-2)} + c.c.
\end{equation}
Here, the higher-spin $\mathcal{N}=2$ Weyl supertensors are defined as in  \cite{Zaigraev:2024ryg}.
At the component level, this action is reduced to the sum of actions with higher derivatives on the Fradkin-Tseytlin fields.
The structure of this action needs the further analysis the results of which will be presented elsewhere.

\medskip

The approaches described in this section are limited  to $\mathcal{N}=2$ multiplets
with {\it integer} highest spins. For constructing $\mathcal{N}=2$ multiplets with the highest {\it half-integer} spin,
it is necessary to consider symmetries of another type. As candidates for such symmetries,  the higher-spin generalizations of ``hidden'' supersymmetry and  $R$-symmetry of
$N=4$ vector multiplet theory can be considered. In support of this proposal, it was shown in ref. \cite{Ivanov:2024bsb} that gauging of hidden supersymmetry and $R$-symmetry
of the hypermultiplet-vector multiplet system \cite{Buchbinder:2020yvf} leads to $\mathcal{N}=2$ superconformal gravitino multiplet.

\section{Consistent superconformal model}
\label{sec: consistent}

An important feature of the superconformal theory is the special structure of operators which generate gauge transformations.
Since the highest term in the operator $\hat{\mathcal{U}}_s$ defined in \eqref{eq: spin s gauge transformations - k-1}
is the most general superfield operator of degree $s$, the commutation relations have the form:
\begin{equation}
    [\hat{\mathcal{U}}_{s_1}, \hat{\mathcal{U}}_{s_2}]
    \sim
    \hat{\mathcal{U}}_{s_1+s_2 - 2}
    +
      \text{lower derivative $\hat{\mathcal{U}}_{s}$ terms}.
\end{equation}
This means that the gauge algebra is closed only provided we involve into game the whole infinite sequence of operators $\hat{\mathcal{U}}_s$.

Thus, in order to build a consistent theory in all orders, it is necessary to consider an infinite tower of $\mathcal{N}=2$ superconformal higher spins:
\begin{equation}
    \hat{\mathbb{H}}^{++} := \sum_{s=1}^{\infty}  \kappa_s \hat{\mathbb{H}}^{++}_{(s)} (J)^{P(s)}.
\end{equation}

As a result, we come to the hypermultiplet action interacting with an infinite tower of $\mathcal{N}=2$ superconformal higher-spin superfields:
    \begin{equation}\label{full vertex}
        S_{full} = - \frac{1}{2} \int d\zeta^{(-4)} \,q^{+a}
            \left(\mathcal{D}^{++} + \hat{\mathbb{H}}^{++} \right) q^+_a.
    \end{equation}
This action is invariant under  the hypermultiplet transformations
\begin{equation}
    \delta_\lambda q^{+a} := - \hat{\mathcal{U}} q^{+a} = - \sum_{s=1}^{\infty} \kappa_s \hat{\mathcal{U}}_{s} q^{+a},
\end{equation}
accompanied by non-Abelian transformations of the gauge superfields:
\begin{equation}\label{eq: NA gauge freedom}
    \delta_\lambda \hat{\mathbb{H}}^{++}
    =
    -
    \left[ \mathcal{D}^{++} + \hat{\mathbb{H}}^{++},   \hat{\mathcal{U}}\right].
\end{equation}
Due to the generic form of the differential operator $\hat{\mathbb{H}}^{++}$, one can ensure the gauge and $\mathcal{N}=2$ superconformal invariances  to all orders,
see details in \cite{Buchbinder:2024pjm}.

 The action \eqref{full vertex}  is a solution for the hypermultiplet theory consistently coupled to an infinite collection of $\mathcal{N}=2$ superconformal gauge higher-spins.
 Emphasize that this result is completely given in terms of unconstrained off-shell harmonic superfields and generalizes the known partial cases for
 conformal interactions of $\mathcal{N}=0$ and $\mathcal{N}=1$ conformal higher spin fields with matter \cite{Segal:2002gd, Bekaert:2010ky, Kuzenko:2022hdv}
 \footnote{After appearance of our work \cite{Buchbinder:2024pjm} the superconformal Lagrangian formulation
 of the hypermultiplet coupled to $\mathcal{N}=2$ gauge higher-spin superfields was also considered in ref. \cite{Kuzenko:2021pqm} (as a continuation of \cite{Kuzenko:2021pqm})
 in another approach  based on the use of $\mathcal{N}=2$ projective superspace.}.

\medskip
\subsubruninhead{$\mathcal{N}=2$ higher-spin superconformal gravity}
\medskip

Using the consistent action \eqref{full vertex} one can build
the quantum effective action
\begin{equation}
    e^{i\Gamma[\hat{\mathbb{H}}^{++}]} = \int \mathcal{D} q^+ \mathcal{D} \tilde{q}^+\; e^{-\frac{i}{2} \int d\zeta^{(-4)} \,q^{+a}
        \left(\mathcal{D}^{++} + \hat{\mathbb{H}}^{++} \right) q^+_a}.
\end{equation}
Since the logarithmically divergent part of such an effective action is
gauge invariant by construction, it can be identified with
$\mathcal{N}=2$ higher-spin superconformal gravity. In order to explicitly construct the
corresponding theory, it is necessary to develop methods
of calculating the effective action for the hypermultiplet in
the background of $\mathcal{N}=2$ higher spins. We expect that at the linearized level such a
theory  reduces to the sum of higher-spin generalizations of Weyl action
and should inevitably contain higher derivatives.

\section{Outlook}
\label{sec: 7}

As another interesting application, we can use the theory \eqref{full vertex}
for a possible construction of the consistent $\mathcal{N}=2$ higher-spin non-conformal supergravity.
As such, we propose the theory of an infinite tower of $\mathcal{N}=2$ superconformal higher spins interacting
with a set of compensating superfields. The compensator set should include  supermultiplets of matter,
$\mathcal{N}=2$ vector multiplet and, presumably, multiplets of half-integer $\mathcal{N}=2$ higher spins.
Such a theory can provide the appropriate arena for realization of the infinite-dimensional superconformal gauge symmetry \eqref{eq: NA gauge freedom}
of higher spins.

As a solution to the equations of motion of this theory, we expect to gain a vacuum which can be identified
with AdS background (see also a recent paper \cite{Didenko:2023txr} on discussion of the higher-spin symmetry breaking).
In this vacuum, $\mathcal{N}=2$ higher-spin superconformal symmetries could be spontaneously broken
to some non-Abelian extension of $\mathcal{N}=2$ higher-spin AdS superalgebra and the theory be reduced to the consistent theory of
interacting AdS $\mathcal{N}=2$ supermultiplets.

To conclude, there remain many interesting open problems on the way to
the completely consistent $\mathcal{N}=2$ higher
spin supergravity. First of all, it is the construction of
various interactions between an infinite tower of the higher spin superfields
and a vector multiplet and the study of the gauge freedom of such an
extended system. Among other open problems we would highlight the
construction of $\mathcal{N}=2$ multiplets of half-integer spins (see ref. \cite{Ivanov:2024bsb}),
as well as of $\mathcal{N}=2$ AdS higher spins and their interactions.
It is worth pointing out that the existence of consistent conformal off-shell interaction of hypermultiplet
to ${\cal N}=2$ superconformal higher spin gauge superfields opens a possibility to develop an approach
to study various aspects of the induced quantum effective action depending on superconformal higher-spin superfields.
All the results attained in these directions would become essential ingredients
of the complete eventual theory.

\medskip
\begin{acknowledgement}
The authors are indebted to Mikhail Vasiliev for numerous interesting discussions. Work of N.Z. was partially supported by the Foundation for the Advancement of Theoretical Physics and Mathematics
``BASIS''.
\end{acknowledgement}
\ethics{Competing Interests}{
The authors have no conflicts of interest to declare that are relevant to the content of this chapter.}

%\eject

\section*{Appendix}
\addcontentsline{toc}{section}{Appendix}
In this appendix, we collect the basic definitions used in the text.

Spinorial notation for the space-time coordinates and derivatives:
\begin{equation}
    x^{\alpha\dot{\alpha}} = \sigma_m^{\alpha\dot{\alpha}} x^m ,
    \qquad
    \partial_{\alpha\dot{\alpha}} = \frac{1}{2}\sigma^m_{\alpha\dot{\alpha}} \partial_m,
    \qquad
    \partial_{\alpha\dot{\alpha}} x^{\beta\dot{\beta}} = \delta^\beta_\alpha \delta_{\dot{\alpha}}^{\dot{\beta}}.
\end{equation}
The rules of tilde conjugation:
\begin{equation}\label{eq: tilde conj}
    \widetilde{x^{\alpha\dot{\alpha}}} = x^{\alpha\dot{\alpha}},
    \quad
    \widetilde{\theta^\pm_\alpha} = \bar{\theta}^\pm_{\dot{\alpha}},
    \quad
    \widetilde{\bar{\theta}^\pm_{\dot{\alpha}}} = - \theta^\pm_\alpha,
    \quad
    \widetilde{u^{\pm i}} = - u_i^\pm,
    \quad
    \widetilde{u^\pm_i} = u^{\pm i},
    \quad
    \widetilde{x^5} = x^5.
\end{equation}
The partial spinor derivatives:
\begin{equation}
    \partial^\pm_{\hat{\alpha}}  = \frac{\partial}{\partial \theta^{\mp \hat{\alpha}}},
    \qquad
    \hat{\alpha} = \{\alpha, \dot{\alpha}\}.
\end{equation}
The multi-index $M$:
\begin{equation}
    M =
    \begin{sqcases}
        \alpha\dot{\alpha}, \alpha+,  \dot{\alpha}+, 5 \quad \quad\text{for non-conformal theories};
        \\
        \alpha\dot{\alpha}, \alpha+,  \dot{\alpha}+, ++ \quad \text{for conformal theories} .
    \end{sqcases}
\end{equation}
The supersymmetry-covariant spinor derivatives:
\begin{equation}\label{eq: spinor cov}
    \begin{split}
        &\mathcal{D}^+_{\hat{\alpha}} := \partial^+_{\hat{\alpha}},
        \quad
        \\&
        \mathcal{D}^-_\alpha
        :=
        -\partial^-_\alpha + 4i \bar{\theta}^{-\dot{\alpha}} \partial_{\alpha\dot{\alpha}}
        -
        2i \theta^-_\alpha \partial_5,
        \quad
        \\
        &
        \bar{\mathcal{D}}^-_{\dot{\alpha}} := - \partial^-_{\dot{\alpha}} - 4i \theta^{-\alpha}
        \partial_{\alpha\dot{\alpha}} - 2i \bar{\theta}^-_{\dot{\alpha}} \partial_5.
    \end{split}
\end{equation}
The Grassmann-parity operator $P(M)$:
\begin{equation}
P(\alpha\dot{\alpha}) = P(++) = 0,\;\;\;
P(\alpha\pm ) =P(\dot{\alpha}\pm )  = 1\,.
\end{equation}
The short-hand notation for the symmetrized spinor indices:
\begin{equation}
    \alpha(s) = (\alpha_1 \alpha_2 \dots \alpha_s),
    \qquad
    \dot{\alpha}(s) = ( \dot{\alpha}_1  \dot{\alpha}_2 \dots    \dot{\alpha}_s).
\end{equation}


\begin{thebibliography}{99}

    \bibitem{St1}
    S.~V.~Ketov and A.~A.~Starobinsky,
    {\it Embedding (R + R2)-Inflation into Supergravity,}
    \href{https://doi.org/10.1103/PhysRevD.83.063512}{Phys. Rev. D \textbf{83} (2011) 063512}, [arXiv:1011.0240 [hep-th]].

    \bibitem{St2}
    S.~V.~Ketov and A.~A.~Starobinsky,
    {\it Inflation and non-minimal scalar-curvature coupling in gravity and supergravity,}
    \href{https://DOI/10.1088/1475-7516/2012/08/022}{JCAP \textbf{08} (2012) 022}, [arXiv:1203.0805
    [hep-th]].

    \bibitem{St3}
    Alexei~A.~Starobinsky, Sergey~V.~Sushkov, Mikhail~S.~Volkov,
    {\it Anisotropy screening in Horndeski cosmologies,}
    \href{https://doi.org/10.1103/PhysRevD.101.064039}{Phys.Rev. D \textbf{101} (2020) 6}, [arXiv:1912.12320 [hep-th]].

    \bibitem{St4}
    Alexey~S.~Koshelev,  K.~Sravan~Kumar, Anupam~Mazumdar, Alexei~A.~Starobinsky,
    {\it Non-Gaussianities and tensor-to-scalar ratio in non-local $R^2$-like inflation,}
    \href{https://doi.org/10.1007/JHEP06(2020)152}{JHEP \textbf{06}, (2020) 152}, [arXiv:2003.00629 [hep-th]].


    %\cite{Vasiliev:1990en}
    \bibitem{Vasiliev:1990en}
    M.~A.~Vasiliev,
    {\it Consistent equation for interacting gauge fields of all spins in (3+1)-dimensions},
    \href{https://doi.org/10.1016/0370-2693(90)91400-6}{Phys. Lett. B \textbf{243} (1990), 378-382}.
    %703 citations counted in INSPIRE as of 10 Dec 2024

    %\cite{Vasiliev:1992av}
    \bibitem{Vasiliev:1992av}
    M.~A.~Vasiliev,
    {\it More on equations of motion for interacting massless fields of all spins in (3+1)-dimensions},
    \href{https://doi.org/10.1016/0370-2693(92)91457-K}{Phys. Lett. B \textbf{285} (1992), 225-234}.
    %440 citations counted in INSPIRE as of 10 Dec 2024

    %\cite{Vasiliev:1999ba}
    \bibitem{Vasiliev:1999ba}
    M.~A.~Vasiliev,
    {\it Higher spin gauge theories: Star product and AdS space},
    \href{https://doi.org/10.1142/9789812793850\_0030}{The Many Faces of the Superworld (2000), 533-610}
    [arXiv:hep-th/9910096 [hep-th]].
    %538 citations counted in INSPIRE as of 10 Dec 2024

    %\cite{Vasiliev:2014vwa}
    \bibitem{Vasiliev:2014vwa}
    M.~A.~Vasiliev,
    {\it Higher-Spin Theory and Space-Time Metamorphoses},
    \href{https://doi.org/10.1007/978-3-319-10070-8\_9}{Lect. Notes Phys. \textbf{892} (2015), 227-264}
    [arXiv:1404.1948 [hep-th]].
    %54 citations counted in INSPIRE as of 10 Dec 2024

    %\cite{Tomasiello:2024jyu}
    \bibitem{Tomasiello:2024jyu}
    A.~Tomasiello,
    {\it Higher spins and Finsler geometry},
    \href{https://doi.org/10.1007/JHEP10(2024)047}{JHEP \textbf{10} (2024), 047}
    [arXiv:2405.00776 [hep-th]].
    %0 citations counted in INSPIRE as of 10 Dec 2024

    %\cite{Bekaert:2004qos}
    \bibitem{Bekaert:2004qos}
    X.~Bekaert, S.~Cnockaert, C.~Iazeolla and M.~A.~Vasiliev,
    {\it Nonlinear higher spin theories in various dimensions},
    [arXiv:hep-th/0503128 [hep-th]].
    %529 citations counted in INSPIRE as of 15 Dec 2024

    %\cite{Bekaert:2010hw}
    \bibitem{Bekaert:2010hw}
    X.~Bekaert, N.~Boulanger and P.~Sundell,
    {\it How higher-spin gravity surpasses the spin two barrier: no-go theorems versus yes-go examples},
    \href{https://doi.org/10.1103/RevModPhys.84.987}{Rev. Mod. Phys. \textbf{84} (2012), 987-1009}
    [arXiv:1007.0435 [hep-th]].
    %305 citations counted in INSPIRE as of 15 Dec 2024

    %\cite{Bekaert:2022poo}
    \bibitem{Bekaert:2022poo}
    X.~Bekaert, N.~Boulanger, A.~Campoleoni, M.~Chiodaroli, D.~Francia, M.~Grigoriev, E.~Sezgin and E.~Skvortsov,
    {\it Snowmass White Paper: Higher Spin Gravity and Higher Spin Symmetry},
    [arXiv:2205.01567 [hep-th]].
    %67 citations counted in INSPIRE as of 15 Dec 2024


    %\cite{Tatarenko:2024csa}
    \bibitem{Tatarenko:2024csa}
    Y.~A.~Tatarenko and M.~A.~Vasiliev,
    {\it Bilinear Fronsdal currents in the AdS$_{4}$ higher-spin theory},
    \href{https://doi.org/10.1007/JHEP07(2024)246}{JHEP \textbf{07} (2024), 246}
    [arXiv:2405.02452 [hep-th]].
    %2 citations counted in INSPIRE as of 10 Dec 2024

    %\cite{Didenko:2024zpd}
    \bibitem{Didenko:2024zpd}
    V.~E.~Didenko and M.~A.~Povarnin,
    {\it All vertices for unconstrained symmetric gauge fields},
    \href{https://doi.org/10.1103/PhysRevD.110.126012}{Phys. Rev. D \textbf{110} (2024) no.12, 126012}
    [arXiv:2409.00808 [hep-th]].
    %0 citations counted in INSPIRE as of 10 Dec 2024

    %\cite{Wess:1992cp}
    \bibitem{Wess:1992cp}
    J.~Wess and J.~Bagger,
    {\it Supersymmetry and supergravity},
    Princeton University Press (1992).
    %461 citations counted in INSPIRE as of 15 Dec 2024

    \bibitem{BK-book}
    I.L. Buchbinder and S.M. Kuzenko,
    {\it Ideas and Methods of Supersymmetry and Supergravity
    or a Walk Through Superspace}, IOP Publishing, Bristol U.K. (1998).

    %\cite{Galperin:1984av}
    \bibitem{Galperin:1984av}
    A.~Galperin, E.~Ivanov, S.~Kalitzin, V.~Ogievetsky and E.~Sokatchev,
    {\it Unconstrained $\mathcal{N}=2$ Matter, Yang-Mills and Supergravity Theories in Harmonic Superspace},
    \href{https://doi.org/10.1088/0264-9381/1/5/004}{Class. Quant. Grav. \textbf{1} (1984), 469-498}
    [erratum: Class. Quant. Grav. \textbf{2} (1985), 127].
%
    %848 citations counted in INSPIRE as of 02 Dec 2024

        \bibitem{18} A.~S.~Galperin, E.~A.~Ivanov, V.~I.~Ogievetsky, E.~S.~Sokatchev,
    {\it Harmonic superspace},
    \href{https://doi.org/10.1017/CBO9780511535109}{Cambridge Monographs on Mathematical
        Physics, Cambridge University Press, 2001, 306 p}.



        %\cite{Buchbinder:2021ite}
        \bibitem{Buchbinder:2021ite}
        I.~Buchbinder, E.~Ivanov and N.~Zaigraev,
        {\it Unconstrained off-shell superfield formulation of $4D,  \mathcal{N}  = 2$ supersymmetric higher spins},
        \href{https://doi.org/10.1007/JHEP12(2021)016}{JHEP \textbf{12} (2021), 016}
        [arXiv:2109.07639 [hep-th]].
        %19 citations counted in INSPIRE as of 03 Dec 2024

        %\cite{Buchbinder:2022kzl}
        \bibitem{Buchbinder:2022kzl}
        I.~Buchbinder, E.~Ivanov and N.~Zaigraev,
        {\it Off-shell cubic hypermultiplet couplings to $ \mathcal{N}  = 2$ higher spin gauge superfields},
        \href{https://doi.org/10.1007/JHEP05(2022)104}{JHEP \textbf{05} (2022), 104}
        [arXiv:2202.08196 [hep-th]].
        %17 citations counted in INSPIRE as of 03 Dec 2024

        %\cite{Buchbinder:2022vra}
        \bibitem{Buchbinder:2022vra}
        I.~Buchbinder, E.~Ivanov and N.~Zaigraev,
        {\it $ \mathcal{N}  = 2$ higher spins: superfield equations of motion, the hypermultiplet supercurrents, and the component structure},
        \href{https://doi.org/10.1007/JHEP03(2023)036}{JHEP \textbf{03} (2023), 036}
        [arXiv:2212.14114 [hep-th]].
        %9 citations counted in INSPIRE as of 03 Dec 2024

        %\cite{Buchbinder:2024pjm}
        \bibitem{Buchbinder:2024pjm}
        I.~Buchbinder, E.~Ivanov and N.~Zaigraev,
        {\it$\mathcal{N}=2$ superconformal higher-spin multiplets and their hypermultiplet couplings},
        \href{https://doi.org/10.1007/JHEP08(2024)120}{JHEP \textbf{08} (2024), 120}
        [arXiv:2404.19016 [hep-th]].
        %4 citations counted in INSPIRE as of 12 Dec 2024


    %\cite{Scherk:1979zr}
    \bibitem{Scherk:1979zr}
    J.~Scherk and J.~H.~Schwarz,
    {\it How to Get Masses from Extra Dimensions},
    \href{https://doi.org/10.1016/0550-3213(79)90592-3}{Nucl. Phys. B \textbf{153} (1979), 61-88}.
    %1308 citations counted in INSPIRE as of 11 Dec 2024

        %\cite{Shaynkman:2001ip}
    \bibitem{Shaynkman:2001ip}
    O.~V.~Shaynkman and M.~A.~Vasiliev,
    {\it Higher spin conformal symmetry for matter fields in (2+1)-dimensions},
    \href{https://doi.org/10.1023/A:1012399417069}{Theor. Math. Phys. \textbf{128} (2001), 1155-1168}
    [arXiv:hep-th/0103208 [hep-th]].
    %54 citations counted in INSPIRE as of 10 Dec 2024

    %\cite{Eastwood:2002su}
    \bibitem{Eastwood:2002su}
    M.~G.~Eastwood,
    {\it Higher symmetries of the Laplacian},
    \href{https://doi.org/10.4007/annals.2005.161.1645}{Annals Math. \textbf{161} (2005), 1645-1665}
    [arXiv:hep-th/0206233 [hep-th]].
    %171 citations counted in INSPIRE as of 10 Dec 2024


    %\cite{Ivanov:2024gjo}
    \bibitem{Ivanov:2024gjo}
    E.~Ivanov and N.~Zaigraev,
    {\it Off-shell invariants of linearized $4D,\mathcal{N}=2$ supergravity in the harmonic approach},
    \href{https://doi.org/10.1103/PhysRevD.110.066020}{Phys. Rev. D \textbf{110} (2024) no.6, 066020}
    [arXiv:2407.08524 [hep-th]].
    %1 citations counted in INSPIRE as of 05 Dec 2024

    %\cite{Fradkin:1979cw}
    \bibitem{Fradkin:1979cw}
    E.~S.~Fradkin and M.~A.~Vasiliev,
    \textit{Minimal set of auxiliary fields in $SO(2)$ extended supergravity},
    %``MINIMAL SET OF AUXILIARY FIELDS IN SO(2) EXTENDED SUPERGRAVITY,''
    \href{https://doi.org/10.1016/0370-2693(79)90774-3}{Phys. Lett. B \textbf{85} (1979) 47-51}.
    %103 citations counted in INSPIRE as of 28 Jan 2024

    %\cite{Fradkin:1979as}
    \bibitem{Fradkin:1979as}
    E.~S.~Fradkin and M.~A.~Vasiliev,
    \textit{Minimal set of auxiliary fields and S-matrix for extended supergravity},
    %``MINIMAL SET OF AUXILIARY FIELDS AND S MATRIX FOR EXTENDED SUPERGRAVITY,''
    \href{https://doi.org/10.1007/BF02776267}{Lett. Nuovo Cim. \textbf{25} (1979) 79-90}.
    %112 citations counted in INSPIRE as of 28 Jan 2024

    %\cite{deWit:1979xpv}
    \bibitem{deWit:1979xpv}
    B.~de Wit and J.~W.~van Holten,
    {\it Multiplets of Linearized $SO(2)$ Supergravity},
    \href{https://doi.org/10.1016/0550-3213(79)90285-2}{Nucl. Phys. B \textbf{155} (1979) 530-542}.
    %191 citations counted in INSPIRE as of 20 Feb 2024

    %\cite{Galperin:1987em}
    \bibitem{Galperin:1987em}
    A.~S.~Galperin, N.~A.~Ky and E.~Sokatchev,
    {\it $\mathcal{N}=2$ Supergravity in Superspace: Solution to the Constraints},
    \href{https://doi.org/10.1088/0264-9381/4/5/022}{Class. Quant. Grav. \textbf{4} (1987), 1235}.
    %55 citations counted in INSPIRE as of 09 Dec 2024

        %\cite{Galperin:1987ek}
    \bibitem{Galperin:1987ek}
    A.~S.~Galperin, E.~A.~Ivanov, V.~I.~Ogievetsky and E.~Sokatchev,
    {\it $\mathcal{N}=2$ Supergravity in Superspace: Different Versions and Matter Couplings},
    \href{https://doi.org/10.1088/0264-9381/4/5/023}{Class. Quant. Grav. \textbf{4} (1987), 1255}.
    %96 citations counted in INSPIRE as of 09 Dec 2024


    %\cite{Ivanov:2022vwc}
    \bibitem{Ivanov:2022vwc}
    E.~Ivanov,
    {\it $\mathcal{N}=2$ Supergravities in Harmonic Superspace},
    \href{https://doi.org/10.1007/978-981-19-3079-9\_43-1}{Handbook of Quantum Gravity,
    2023,
    Springer Nature Singapore,
    1-50}
    [arXiv:2212.07925 [hep-th]].
    %5 citations counted in INSPIRE as of 09 Dec 2024



    %\cite{Zupnik:1998td}
    \bibitem{Zupnik:1998td}
    B.~M.~Zupnik,
    {Background harmonic superfields in $\mathcal{N}=2$ supergravity},
    \href{https://doi.org/10.1007/BF02557138}{Theor. Math. Phys. \textbf{116} (1998), 964-977}
    [arXiv:hep-th/9803202 [hep-th]].
    %10 citations counted in INSPIRE as of 09 Dec 2024

    %\cite{Freedman:2012zz}
    \bibitem{Freedman:2012zz}
    D.~Z.~Freedman and A.~Van Proeyen,
    {\it Supergravity},
    \href{https://doi.org/10.1017/CBO9781139026833}{Cambridge University Press, 2012, 607p}.
%   ISBN 978-1-139-36806-3, 978-0-521-19401-3

    %251 citations counted in INSPIRE as of 12 Dec 2024



    %\cite{deWit:1979dzm}
    \bibitem{deWit:1979dzm}
    B.~de Wit, J.~W.~van Holten and A.~Van Proeyen,
    {\it Transformation Rules of $\mathcal{N}=2$ Supergravity Multiplets},
    \href{https://doi.org/10.1016/0550-3213(80)90125-X}{Nucl. Phys. B \textbf{167} (1980), 186}.
    %243 citations counted in INSPIRE as of 17 Dec 2024

    %\cite{Bergshoeff:1980is}
    \bibitem{Bergshoeff:1980is}
    E.~Bergshoeff, M.~de Roo and B.~de Wit,
    {\it Extended Conformal Supergravity},
    \href{https://doi.org/10.1016/0550-3213(81)90465-X}{Nucl. Phys. B \textbf{182} (1981), 173-204}.
    %350 citations counted in INSPIRE as of 09 Dec 2024

    %\cite{deWit:1980lyi}
    \bibitem{deWit:1980lyi}
    B.~de Wit, J.~W.~van Holten and A.~Van Proeyen,
    {\it Structure of $\mathcal{N}=2$ Supergravity},
    \href{https://doi.org/10.1016/0550-3213(81)90211-X}{Nucl. Phys. B \textbf{184} (1981), 77}
    [erratum: \href{https://doi.org/10.1016/0550-3213(83)90548-5}{Nucl. Phys. B \textbf{222} (1983), 516}].
    %271 citations counted in INSPIRE as of 10 Dec 2024

    %\cite{Kuzenko:2015jxa}
    \bibitem{Kuzenko:2015jxa}
    S.~M.~Kuzenko and J.~Novak,
    {\it On curvature squared terms in $\mathcal{N}=2$ supergravity},
    \href{https://doi.org/10.1103/PhysRevD.92.085033}{Phys. Rev. D \textbf{92} (2015) no.8, 085033}
    [arXiv:1507.04922 [hep-th]].
    %20 citations counted in INSPIRE as of 17 Dec 2024

    %\cite{Kuzenko:2022ajd}
    \bibitem{Kuzenko:2022ajd}
    S.~M.~Kuzenko, E.~S.~N.~Raptakis and G.~Tartaglino-Mazzucchelli,
    {\it Covariant Superspace Approaches to $\mathscr {N}=\text{2}$ Supergravity},
    \href{https://doi.org/10.1007/978-981-19-3079-9\_44-1}{Handbook of Quantum Gravity (2023),
    publisher="Springer Nature Singapore", 1-61}
    [arXiv:2211.11162 [hep-th]].
    %18 citations counted in INSPIRE as of 17 Dec 2024


        %\cite{Galperin:1985tn}
    \bibitem{Galperin:1985tn}
    A.~Galperin, E.~Ivanov and V.~Ogievetsky,
    {\it Superspace Actions and Duality Transformations for $\mathcal{N}=2$ Tensor Multiplets},
    Sov. J. Nucl. Phys. \textbf{45} (1987), 157, \href{https://doi.org/10.1088/0031-8949/1987/T15/025}{Phys.Scripta T 15 (1987) 176}.
    %35 citations counted in INSPIRE as of 09 Dec 2024



        %\cite{Zaigraev:2024ryg}
    \bibitem{Zaigraev:2024ryg}
    N.~Zaigraev,
    {\it $\mathcal{N}=2$ higher-spin supercurrents},
    \href{https://doi.org/10.1016/j.physletb.2024.139056}{Phys. Lett. B \textbf{858} (2024), 139056}
    [arXiv:2408.00668 [hep-th]].
    %0 citations counted in INSPIRE as of 13 Dec 2024


    %\cite{Fradkin:1985am}
    \bibitem{Fradkin:1985am}
    E.~S.~Fradkin and A.~A.~Tseytlin,
    {\it Conformal supergravity},
    \href{https://doi.org/10.1016/0370-1573(85)90138-3}{Phys. Rept. \textbf{119} (1985), 233-362}.
    %451 citations counted in INSPIRE as of 13 Dec 2024

        %\cite{Segal:2002gd}
    \bibitem{Segal:2002gd}
    A.~Y.~Segal,
    {\it Conformal higher spin theory},
    \href{https://doi.org/10.1016/S0550-3213(03)00368-7}{Nucl. Phys. B \textbf{664} (2003), 59-130}
    [arXiv:hep-th/0207212 [hep-th]].
    %192 citations counted in INSPIRE as of 10 Dec 2024

     %\cite{Ivanov:2024bsb}
     \bibitem{Ivanov:2024bsb}
     E.~Ivanov and N.~Zaigraev,
     \textit{N=2 superconformal gravitino in harmonic superspace},
     \href{https://doi.org/10.1016/j.physletb.2025.139333}{Phys. Lett. B \textbf{862} (2025), 139333}
     [arXiv:2412.14822 [hep-th]].
     %0 citations counted in INSPIRE as of 27 Feb 2025

    %\cite{Buchbinder:2020yvf}
    \bibitem{Buchbinder:2020yvf}
    I.~L.~Buchbinder, E.~A.~Ivanov and V.~A.~Ivanovskiy,
    {\it Superfield realization of hidden $R$-symmetry in extended supersymmetric gauge theories and its applications},
    \href{https://doi.org/10.1007/JHEP04(2020)126}{JHEP \textbf{04} (2020), 126}
    [arXiv:2001.01649 [hep-th]].
    %1 citations counted in INSPIRE as of 17 Dec 2024



    %\cite{Bekaert:2010ky}
    \bibitem{Bekaert:2010ky}
    X.~Bekaert, E.~Joung and J.~Mourad,
    {\it Effective action in a higher-spin background},
    \href{https://doi.org/10.1007/JHEP02(2011)048}{JHEP \textbf{02} (2011), 048}
    [arXiv:1012.2103 [hep-th]].
    %129 citations counted in INSPIRE as of 10 Dec 2024

    %\cite{Kuzenko:2022hdv}
    \bibitem{Kuzenko:2022hdv}
    S.~M.~Kuzenko, M.~Ponds and E.~S.~N.~Raptakis,
    {\it Conformal Interactions Between Matter and Higher-Spin (Super)Fields},
    \href{https://doi.org/10.1002/prop.202200157}{Fortsch. Phys. \textbf{71} (2023) no.1, 1}
    [arXiv:2208.07783 [hep-th]].
    %16 citations counted in INSPIRE as of 10 Dec 2024

    %\cite{Kuzenko:2024vms}
    \bibitem{Kuzenko:2024vms}
    S.~M.~Kuzenko and E.~S.~N.~Raptakis,
    {\it Towards $\mathcal{N}=2$ superconformal higher-spin theory},
    \href{https://doi.org/10.1007/JHEP11(2024)013}{JHEP \textbf{11} (2024), 013}
    [arXiv:2407.21573 [hep-th]].
    %1 citations counted in INSPIRE as of 13 Dec 2024

    %\cite{Kuzenko:2021pqm}
    \bibitem{Kuzenko:2021pqm}
    S.~M.~Kuzenko and E.~S.~N.~Raptakis,
    {\it Extended superconformal higher-spin gauge theories in four dimensions},
    \href{https://doi.org/10.1007/JHEP12(2021)210}{JHEP \textbf{12} (2021), 210}
    [arXiv:2104.10416 [hep-th]].
    %21 citations counted in INSPIRE as of 17 Dec 2024


    %\cite{Didenko:2023txr}
    \bibitem{Didenko:2023txr}
    V.~E.~Didenko and A.~V.~Korybut,
    {\it Toward higher-spin symmetry breaking in the bulk},
    \href{https://doi.org/10.1103/PhysRevD.110.026007}{Phys. Rev. D \textbf{110} (2024) no.2, 026007}
    [arXiv:2312.11096 [hep-th]].
    %3 citations counted in INSPIRE as of 10 Dec 2024







\end{thebibliography}
\end{document}